\documentclass{aa}

\usepackage{graphicx}
\usepackage{placeins}
\usepackage{txfonts}
\usepackage{natbib}
\usepackage{subfigure}
\begin{document}

   \title{The disc origin of the Milky Way bulge:}

   \subtitle{Dissecting the chemo-morphological relations using N-body simulations and APOGEE}
%F. Fragkoudi
%          \inst{1}\fnmsep\thanks{francesca.fragkoudi@obspm.fr}
   \author{F. Fragkoudi
          \inst{1,2}
          \and
          P. Di Matteo\inst{2}
           \and
          M. Haywood\inst{2} 
           \and
           M. Schultheis\inst{3}
          \and
          S. Khoperskov\inst{2}
          \and
          A. G\'{o}mez\inst{2} 
          \and
          F. Combes\inst{4,5} 
          }

   \institute{Max-Planck-Institut f\"{u}r Astrophysik, Karl-Schwarzschild-Str. 1, 85741 Garching, Germany\\
            \email{ffrag@mpa-garching.mpg.de}
        \and
         GEPI, Observatoire de Paris, PSL University, CNRS, Place Jules Janssen, 92195, Meudon, France
          \and
             Laboratoire Lagrange, Universit\'{e} C\^{o}te d'Azur, Observatoire de la C\^{o}te d'Azur, CNRS, Bd de l'Observatoire, 06304 Nice, France
         \and
             Observatoire de Paris, LERMA, CNRS, PSL Univ., UPMC, Sorbonne Univ., F-75014, Paris, France
         \and
             College de France, 11 Place Marcelin Berthelot, 75005, Paris, France
             }

   \date{}

% \abstract{}{}{}{}{} 
% 5 {} token are mandatory
 
  \abstract
   {There is a long-standing debate on the origin of the metal-poor stellar populations of the Milky Way (MW) bulge, with the two leading scenarios being that these populations are either \emph{i}) part of a classical metal-poor spheroid or \emph{ii}) the same population as the chemically defined thick disc seen at the Solar neighbourhood. Here we test whether the latter scenario can reproduce the observed chemical properties of the MW bulge. To do so we compare an N-body simulation of a composite (thin+thick) stellar disc -- which evolves secularly to form a bar and a boxy/peanut (b/p) bulge -- to data from APOGEE DR13. This model, in which the thick disc is massive and centrally concentrated, can reproduce the morphology of the metal-rich and metal-poor stellar populations in the bulge, as well as the mean metallicity and [$\alpha$/Fe] maps as obtained from the APOGEE data. It also reproduces the trends, in both longitude and latitude, of the bulge metallicity distribution function (MDF). Additionally, we show that the model predicts small but measurable azimuthal metallicity variations in the inner disc due to the differential mapping of the thin and thick disc in the bar. We therefore see that the chemo-morphological relations of stellar populations in the MW bulge are naturally reproduced by mapping the thin and thick discs of the inner MW into a b/p. }


   \maketitle
%
%-------------------------------------------------------------------

\section{Introduction}

%overview, cosmological interest. History/context - change of paradigm
The formation mechanism of the Milky Way (MW) bulge has been a topic of debate in the last few decades, as it has implications on the overall formation history of the Galaxy (e.g. \citealt{Caluraetal2012, Obrejaetal2013, Tisseraetal2018, Grandetal2018}) and indeed on our understanding of galaxy formation and evolution in general. 
The recent cataclysm of data from high resolution spectroscopic surveys (e.g. \citealt{ Bovyetal2012, Kunderetal2012, Freemanetal2013, RojasArriagadaetal2014, Majewskietal2017, Zoccalietal2017}) has lead to new insights on the morphological, kinematic and chemical properties of the MW bulge and disc. These have lead to a paradigm shift in terms of the bulge's origin, with recent works showing that the MW bulge might be the result of disc instabilities \citep{Shenetal2010,BekkiTsujimoto2011,Nessetal2012,WeggGerhard2013,DiMatteoetal2015,Portailetal2017}, rather than being a dispersion dominated spheroid formed via, for example, dissipational collapse (e.g. \citealt{Eggenetal1962}).

%morphology: link between bulge and disc instabilities
The link between disc instabilities and the MW bulge becomes clear when examining the bulge's morphology, namely its characteristic X-shape, seen from images in the near- and mid-infrared \citep{Dweketal1995,NessLang2016}. This X-shape has been interpreted as being due to the presence of a boxy/peanut (b/p) bulge, a structure which forms due to the vertical heating of stellar bars through resonances and/or the buckling instability \citep{CombesSanders1981,Combesetal1990,Rahaetal1991, Athanassoula2005,MartinezValpuestaetal2006,Quillenetal2014}. The b/p morphology is also evidenced by a split in the red clump magnitude distribution along the bulge minor axis, where two peaks are seen, one on the near and one on the far side of the galactic centre, indicating that these lines of sight are crossing the ``arms'' of the X-shaped bulge \citep{Natafetal2010, McWilliamandZoccali2010}.

%MDF of bulge and relation between morphology and metallicity
While the morphology of the bulge points to a secular disc origin, the chemistry of the bulge has proven more difficult to disentangle. Its metallicity distribution function (MDF) is broad \citep{McWilliamRich1994,Hilletal2011,Nessetal2013a, RojasArriagadaetal2014,Zoccalietal2017,GarciaPerrezetal2017}, pointing towards the co-existence of multiple stellar populations; while it is generally agreed that the metal-rich populations have a disc origin, debate still remains over the origin of the metal-poor populations, which make up a large fraction of the mass of the MW bulge. Interestingly, there is a clear relation between the metallicity and morphology of stellar populations in the bulge. For example, the fractional contribution of metal-poor populations in the bulge increases with increasing distance from the galactic plane, which gives rise to a vertical metallicity gradient in the bulge \citep{Minnitietal1995,Zoccalietal2008,Nessetal2013a,Gonzalezetal2013}. 
The relation between metallicity and morphology is also evident in the shape of the b/p bulge of the MW; the split in the red clump distribution is seen more prominently in the kinematically coldest and most metal-rich populations ([Fe/H] > 0) compared to warmer and more metal-poor populations ([Fe/H] < 0), where the split appears weaker and at larger heights from the plane, or for the most metal-poor stars, does not appear at all (e.g. \citealt{Nessetal2013a,RojasArriagadaetal2014}). This has been interpreted as evidence for two different components in the MW bulge, a metal-rich ([Fe/H] > 0) component with a disc origin, and a metal-poor ([Fe/H] < 0), classical spheroid component (e.g. \citealt{Zoccalietal2008,RojasArriagadaetal2014}). On the other hand, recent studies using N-body models with multiple disc populations have shown that this behaviour can also be explained within a pure disc context, where the bulge is made up of thin and thick disc stars; in such a scenario, the morphology of the b/p will depend on the kinematics of each population  \citep{DiMatteo2016, Athanassoulaetal2017,Debattistaetal2017,Fragkoudietal2017}.

%connection between bulge and disc: importance of thick disc, summarise things we know about thick disc of MW
If the MW bulge formed through disc instabilities there will of course be an intimate link between the stellar populations of the inner disc and bulge (for a detailed discussion on the link between the MW bulge and disc we refer the reader to the review by \citealt{DiMatteo2016}). It has been shown that the MW's disc has a continuum of mono-abundance stellar populations \citep{Bovyetal2012}, of which the most metal-poor and $\alpha$-enhanced have shorter scale lengths and larger scale heights \citep{Bensbyetal2011,Bovyetal2012,Bensbyetal2014,Bovyetal2016,Mackerethetal2017}. These centrally concentrated and $\alpha$-enhanced populations are referred to as the chemically defined thick disc\footnote{This is in order to distinguish from the geometrically defined thick disc. For a discussion on this see \cite{Minchevetal2015}.}, which we will refer to simply as the thick disc in what follows.
The stellar populations of the thick disc exhibit a tight age-metallicity relation at the Solar vicinity (\citealt{Haywoodetal2013}) hinting at an in-situ formation in a well mixed ISM which likely arose due to turbulent processes at high redshift \citep{Lehnertetal2014}. Such processes could arise, for example, due to gas rich minor mergers or intense gas accretion leading to a turbulent and clumpy ISM \citep{Brooketal2004,Bournaudetal2009} -- which in turn leads to the upside-down formation of disc galaxies \citep{Birdetal2013,Martigetal2014,Grandetal2016}. Recent studies have also suggested that the thick disc of the MW is at least as massive as the thin disc \citep{Snaithetal2014,Haywoodetal2015} which further hints to the importance that this population will have in the central regions of the MW, due both to its large mass and short scale length.

%What do we do here, what is the model?
We explore the consequences of a pure (thin+thick) disc origin for the MW bulge, in a series of papers, by utilising N-body simulations in which the bulge forms out of the secular evolution of a composite disc -- made of a thin and massive and centrally concentrated thick disc: In Di Matteo et al. (subm.) we explore the necessity of a thick disc in the bulge formation process, while in Di Matteo et al. (in prep.) we explore the relation between the morphology and kinematics of the MW bulge. 
In this paper, we explore the relation between morphology and chemical abundances of stellar populations in the MW bulge; the model evolves in isolation and forms a bar, which subsequently maps the thin and thick discs into a b/p bulge. We compare the model to data of the MW bulge and inner disc from the near-infrared spectroscopic survey APOGEE \citep{Majewskietal2017}. As we will show, the model reproduces well all the explored trends, thus hinting at the possible pure disc origin of the MW bulge, in which the metal-poor and $\alpha$-enhanced populations seen in the bulge are simply the thick disc of the MW, being mapped into the inner regions by the bar.

%In this scenario, the metal-poor populations in the bulge have their origin both in rapid and violent processes, as well as in secular processes; while the b/p morphology arises through disc instabilities, the stellar populations that belong to the thick disc, and which are part of the MW bulge, were born at high redshifts when the galaxy was forming stars with a high SFR \citep{Lehnertetal2014}.

%outline of paper
The outline of the paper is as follows: in Section \ref{sec:model} we describe the N-body simulation as well as the forward modelling applied in order to compare it to APOGEE data of the MW bulge. In Section \ref{sec:morph} we examine the morphology of the inner disc stellar populations after they are mapped into the b/p bulge. In Section \ref{sec:metal} we show what the implications of this mapping will be for the abundances in the MW bulge, by exploring mean metallicity and [$\alpha$/Fe] maps as well as the MDF of the bulge. In Section \ref{sec:azimuth} we explore the azimuthal metallicity variations which arise in the inner disc when viewing the model face-on. In Section \ref{sec:discussion} we discuss some of the most important points raised by this study as well as the limitations of the model, and in Section \ref{sec:summary} we finish with the conclusions of this study.

 %----------------------------------------------------------------------------------------
%	SECTION models
%----------------------------------------------------------------------------------------
%--------------------------------------------------------------------
\section{The model}
\label{sec:model}

\begin{figure}
\centering
\includegraphics[width=0.95\linewidth]{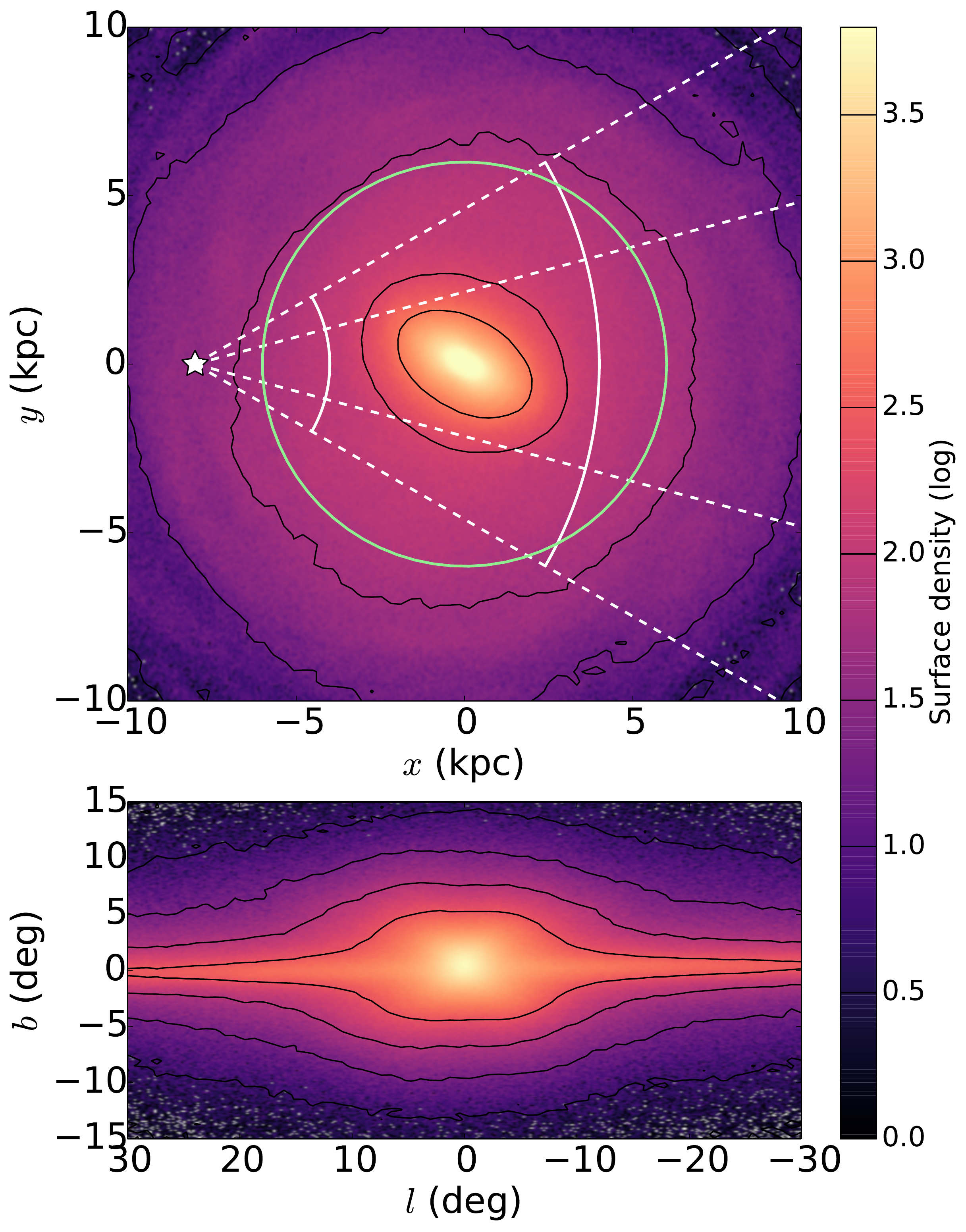}
\caption{\emph{Top panel:} Surface density of the model in the $xy$ plane in arbitrary units. The white star indicates the position of the Sun. The bar semi-major axis has an angle of 30 degrees with respect to the Sun-Galactic centre line. The dashed lines indicate $l$=$\pm$15 and $l$=$\pm$30 angles. The solid white lines indicate 4 and 12\,kpc distance from the Sun. The green circle delineates  a radius of 6\,kpc, what we refer to as the inner disc of the model. \emph{Bottom panel:} Edge-on view of the model in galactic longitude $l$ and latitude $b$.}
\label{fig:model1}
\end{figure}

We explore a purely collisionless N-body simulation of a composite disc galaxy with stellar mass and rotation curve compatible to those of the Milky Way. The vertically continuous stellar populations seen in the Milky Way disc (see \citealt{Bovyetal2012}, Figure 5) are discretised into three co-spatial discs, which can roughly be associated -- morphologically, kinematically and chemically -- to the metal-rich thin disc, the young thick disc and the old thick disc seen at the solar vicinity (nomenclature as in \citealt{Haywoodetal2013}). This composite stellar disc is then embedded in a live dark matter halo and let to evolve in isolation. After $\sim$1\,Gyr a strong stellar bar forms which transfers angular momentum from the inner regions of the system to the outer disc and dark matter halo, thus bringing the model into a lower energy configuration state \citep{LyndenBellKalnajs1972}. After about 5\,Gyrs of evolution a prominent boxy/peanut bulge forms due to vertical instabilities in the bar. In what follows we explore one of the final snapshots of the simulation, after it has evolved for 7\,Gyrs, once it has developed a bar and a boxy/peanut bulge. 

As we would like to compare directly to observables of the Milky Way bulge, we ``observe'' the simulation from the Sun's location within the disc, i.e. at a distance of 8\,kpc from the galactic centre, with the bar rotated at 30 degrees with respect to the galactocentric line-of-sight \citep{BlandHawthornGerhard2016}. In Figure \ref{fig:model1} we show the surface density of the model, face-on in $xy$ (top panel) and edge-on in Galactic longitude and latitude ($l,b$, bottom panel). In the top panel of the figure, the position of the Sun is indicated with the white star, and we indicate longitudes of $\pm$15 and $\pm$30 degrees with the dashed lines. The inner disc of the model ($R$ < 6\,kpc), i.e. the region on which we focus on in this work, is delineated with the light green line. It is non-trivial to define a strict separation between the bulge and the inner disc, since, in our model, the bulge is a b/p formed out of inner disc stars, rather than a well separated component. In what follows, we will refer to the bulge as the particles within |$l$, $b$| < 10\,deg, with distances between 4 and 12\,kpc from the Sun (solid white lines), which is a commonly used definition in Galactic studies of the bulge. When we refer to the inner disc we are referring to everything between |$l$| = 10-30\,deg, which roughly corresponds to the disc inside 6\,kpc as can be seen in Figure \ref{fig:model1}. 

In the next subsections we describe the N-body simulation and its initial conditions in more detail, and we show how we take some of the observational biases into account when comparing our model to data of the MW bulge.

\subsection{N-body simulation}
\label{sec:5models}
%--------------------------------------
\begin{table*}
\centering
\label{tab:info}
\begin{tabular}{ l r | c | c | c | c | c | c | c | c } 
& & $r_D$ (kpc) & $h_z$ (kpc)  & $M$ ($M_{\odot}$) & $n_p$  & [Fe/H] (dex) & $\sigma_{[Fe/H]}$ (dex) & [$\alpha$/Fe] (dex) & $\sigma_{[\alpha/Fe]}$ \\ \hline
&\emph{Cold (D1)} & 4.8 & 0.15 & 4.21 $\times$ $10^{10}$ & 5000000  & 0.25 & 0.15 & 0.09 & 0.04 \\ \hline
&\emph{Intermediate (D2)} & 2 & 0.3 & 2.57 $\times$ $10^{10}$ & 3000000 & -0.26 & 0.2 & 0.15 & 0.05 \\ \hline
&\emph{Hot (D3)} & 2 & 0.6 & 1.86 $\times$ $10^{10}$ & 2000000  & -0.62 & 0.26 & 0.22 & 0.04 \\ \hline
%& \emph{DM Halo} & 3 & 0.6 & 3.68 $\times$ $10^{11}$ & 5000000  & - & - \\ \hline
\end{tabular}
\vspace{0.1cm}
\caption{Properties of the simulations used in this study. From left to right: the characteristic radius of the population, the characteristic height of the population, mass of the component, number of particles in component, the mean metallicity and dispersion in metallicity, and the mean [$\alpha$/Fe] and dispersion in [$\alpha$/Fe].}
\end{table*}

\begin{figure}
\centering
\includegraphics[width=0.9\linewidth]{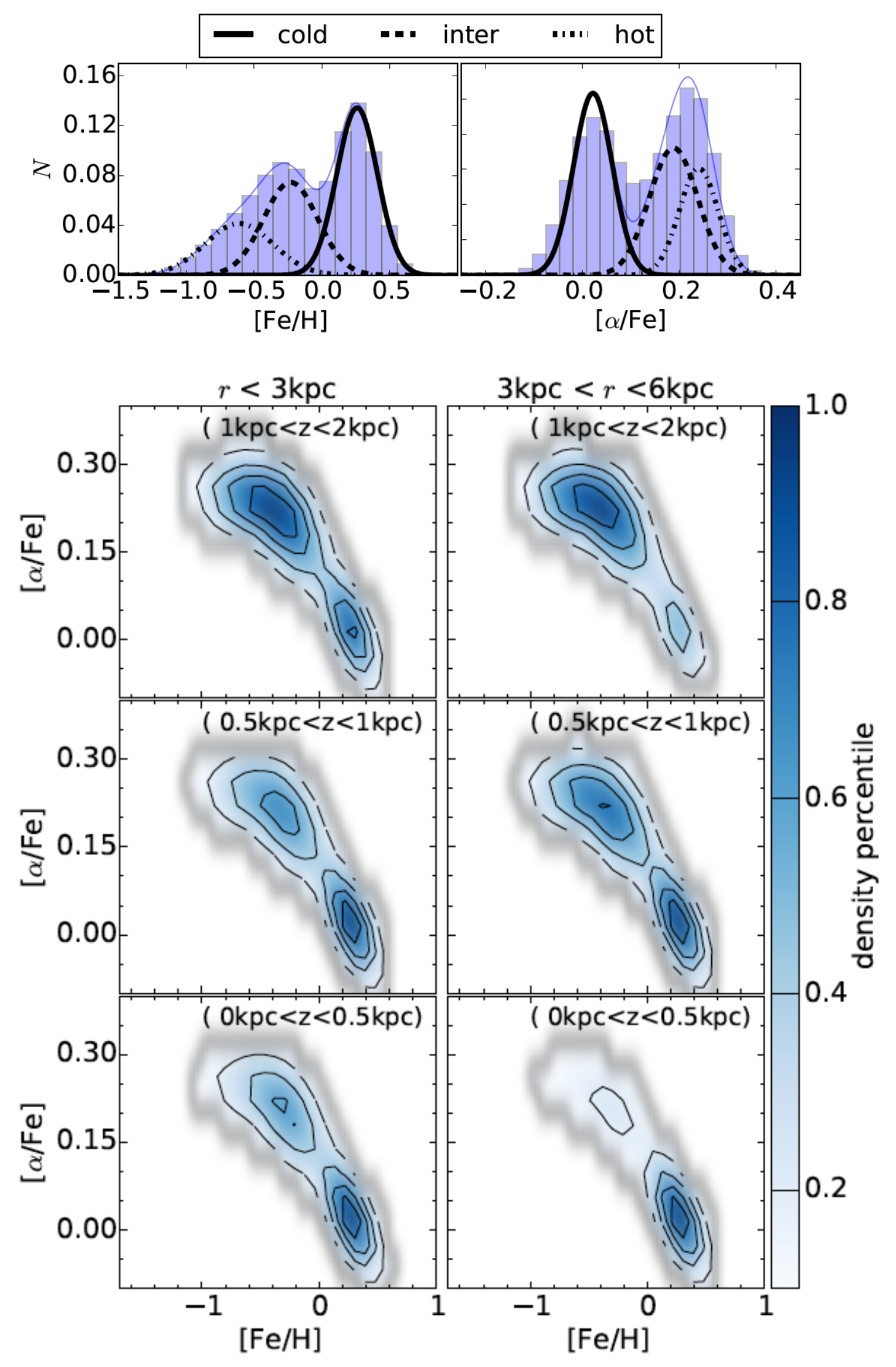}
\caption{\emph{Top panel:} The MDF (left) and ADF (right) of the inner 6\,kpc of the model. The curves correspond to the gaussians assigned to the cold disc (solid), the intermediate disc (dashed) and hot disc (dashed-dotted). \emph{Bottom panels:} We show the [$\alpha$/Fe] vs [Fe/H] plane for our model inside 6\,kpc, separating into particles in the inner 3\,kpc and those between 3 < $r$ < 6\,kpc. We also separate into three vertical bins as indicated at the top of each panel. The gray colour corresponds to values below 5\%.}
\label{fig:discsmdfadf}
\end{figure}

\begin{figure}
\centering
\includegraphics[width=0.9\linewidth]{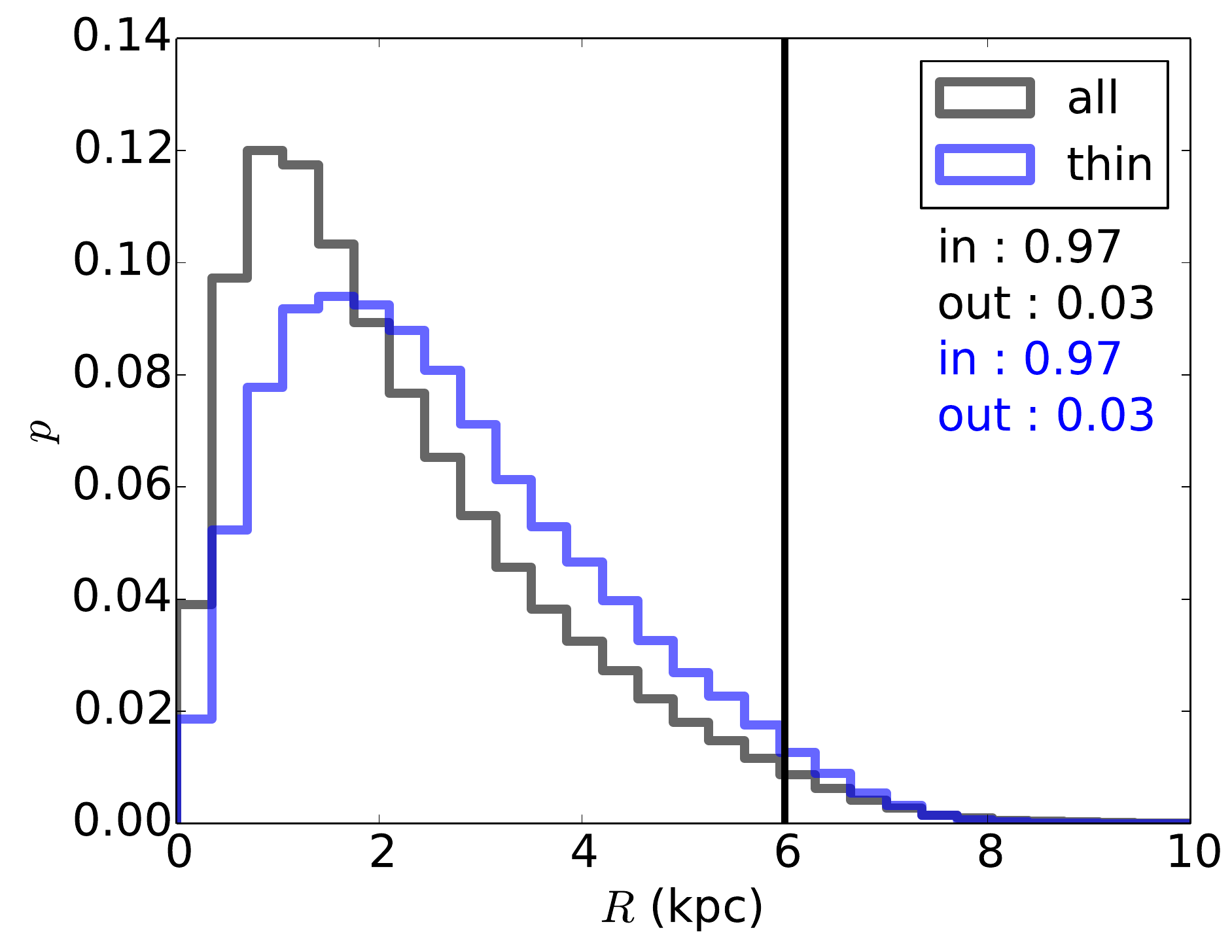}
\caption{The initial distribution of stars that end up within 6\,kpc from the galactic centre after 7\,Gyr of evolution. The stars within 6\,kpc are what we loosely term the inner disc of the Milky Way. We see that almost all stars originate from within 6\,kpc, with about 3\% contamination from the ``outer'' disc.}
\label{fig:in6kpc}
\end{figure}

The three disc components which make up our composite stellar disc correspond to a kinematically cold and metal-rich ([Fe/H] > 0) thin disc (\emph{D1}), to an intermediate disc (\emph{D2}) with intermediate kinematics and metallicities (0 > [Fe/H] > -0.5), and to a kinematically hot and metal-poor (-0.5 > [Fe/H] > -1) disc (\emph{D3}). The intermediate and hot discs (\emph{D2} and \emph{D3}) correspond to the thick disc of the MW, and represent the young and old thick disc respectively, seen at the Solar vicinity, while the cold disc corresponds to the thin disc of the MW (nomenclature as in \citealt{Haywoodetal2013}). The intermediate and hot discs have a combined mass of 50\% of the total stellar mass of the model (in agreement with the mass growth of the MW disc as estimated by \cite{Snaithetal2014,Snaithetal2015}), and both have shorter scalelengths than the cold disc, thus making these populations more concentrated in the central regions of the disc. For a summary of their properties we refer the reader to Table \ref{tab:info}. We interchangeably refer to these three disc populations as either \emph{D1}, \emph{D2} and \emph{D3} or cold, intermediate and hot, while the thin disc refers to disc \emph{D1} and the thick disc refers to both \emph{D2} and \emph{D3}.

In what follows we are only interested in the \emph{inner} disc, since the outer disc does not participate in the b/p bulge (see \citealt{DiMatteoetal2014} and \citealt{Halleetal2015}). The inner disc ($r$ $\sim$ 6\,kpc) roughly corresponds to the location of the Outer Lindblad Resonance (OLR) in our model, which is located at  $\sim$7\,kpc. 
The fact that only the inner disc will contribute to the bulge region, and that ``contamination'' from the outer disc is negligible can be seen in Figure \ref{fig:in6kpc}, where we show the initial radial distribution of stars that end-up within 6\,kpc at the end of the simulation; we see that only a 3\% percent of stars originate from outside 6\,kpc. Therefore the stars that make up the MW bulge all have an origin from the inner disc. 

The particles in the three discs are assigned a metallicity ([Fe/H]) and $\alpha$-abundance ([$\alpha$/Fe]) by drawing randomly from normal distributions, where each disc has a mean metallicity and dispersion, and a mean $\alpha$-abundance and dispersion (see table \ref{tab:info}). The metallicities and $\alpha$-abundances are assigned such that we can reproduce the [$\alpha$/Fe] vs. [Fe/H] plane of the inner Milky Way disc as obtained from APOGEE data (see e.g. \citealt{Haywoodetal2016}).
In the top panel of Figure \ref{fig:discsmdfadf} we show the metallicity distribution function (MDF) and the $\alpha$ distribution function (ADF) of the model inside $\sim$6\,kpc, where the three gaussians correspond to the metallicity and $\alpha$ values that we assign to each disc. We point out that although we have three discs in our model, we only have two peaks in the MDF, with a dip at [Fe/H]$\sim$0. In what follows, when we refer to the metal-rich (MR) population ([Fe/H] > 0) we refer to the cold disc population of our model, while when we refer to the metal-poor (MP) population we refer to both the intermediate and hot disc populations (which have [Fe/H] < 0).

In the bottom panels of Figure \ref{fig:discsmdfadf} we show the [$\alpha$/Fe] vs. [Fe/H] plane for the inner 6\,kpc of our model. We separate into the innermost 3\,kpc, and between 3<$r$<6\,kpc and take three different cuts with height above the plane. We see that the alpha-enhanced metal-poor population is more dominant at higher latitudes and decreases as we move closer to the plane, while the metal-rich alpha-poor population increases close to the plane. These overall trends are similar to what is seen in the MW (see for example Figure 6 of \citealt{NessFreeman2016} and \citealt{Haydenetal2015}).

We point out that discs \emph{D2} and \emph{D3} have a flat radial metallicity gradient, since we assume that the ISM was well-mixed at the time the young and old thick disc formed ($z$\,>\,1), due to a high star formation rate (>\,10 $M_{\odot}$/yr), meaning that  the galaxy was likely in a bursty and turbulent state (\citealt{Lehnertetal2014,Wuytsetal2016, Maetal2017}). The cold disc in our model is associated to the final more quiescent phase of the Milky Way in the last $\sim$7-8 Gyr (see \citealt{Snaithetal2014,Haywoodetal2016}). For simplicity we also assume a flat metallicity gradient for the inner thin disc. More complex gradients -- and their evolution with time due to radial migration -- will be the subject of future work. 
While it is possible that the inner thin disc could have a negative gradient -- and the thin disc over the whole extent of the MW disc indeed has a negative gradient (e.g. \citealt{Andersetal2017}) -- we point out that the overall gradient in the disc could not have been very steep at high redshifts since this would not reproduce the APOGEE metallicity maps \citep{Fragkoudietal2017b}. 
In terms of the vertical gradients, due to the fact that \emph{D2} and \emph{D3} have a larger scaleheight than the thin disc, this leads to a global negative vertical metallicity gradient at the start of the simulation. 

Lastly, we emphasise that our model is meant to be representative of the \emph{inner} Milky Way (i.e. up to the OLR -- see e.g. \citealt{Dehnen2000} but also \citealt{PerezVillegasetal2017}) and is evolved over secular timescales, i.e. 7\,Gyr until a bar and b/p form. The snapshot we analyse is re-scaled so that the bar has a length of $\sim$4.5\,kpc, similar to the length of the MW bar (e.g. \citealt{BlandHawthornGerhard2016}).

\subsection{Initial Conditions \& code}

The simulation is evolved self-consistently, in isolation, from an initial axisymmetric configuration in equilibrium. The initial conditions of the model are obtained using the algorithm of \cite{Rodionovetal2009}, the so-called ``iterative'' method.
The algorithm constructs equilibrium phase models for stellar systems, thus avoiding the problem of the initial relaxation process often observed in N-body models of discs. This is achieved using a constrained evolution, so that the equilibrium solution has a number of desired parameters. In our case we impose the density distributions of the discs, which are described by a Miyamoto-Nagai profile \citep{BT2008}, where each disc has a characteristic radius $r_D$ given in Table \ref{tab:info}. The velocity dispersion is let to evolve unconstrained, with the requirement that the initial conditions (ICs) generated are in equilibrium. 

The time integration algorithm used is a recently developed parallel MPI Tree-code which takes into account the adaptive spatial decomposition of particle space between nodes. The multi-node Tree-code is based on the 256-bit AVX instructions which significantly speed up the floating point vector operations and sorting algorithms (Khoperskov et al. in prep). In total we employ 15 million particles in the model, 10 million in the disc and five in the dark matter halo. The tolerance parameter of the tree-code is $\theta$ = 0.7, the time step is $\Delta t$ = 2 $\times$ 10$^5$ and for smoothing we use a Plummer potential with $\epsilon$ = 50\,pc.

\subsection{Comparing with APOGEE DR13 data}
\label{sec:compare}

\begin{figure*}
\centering
\includegraphics[width=0.9\linewidth]{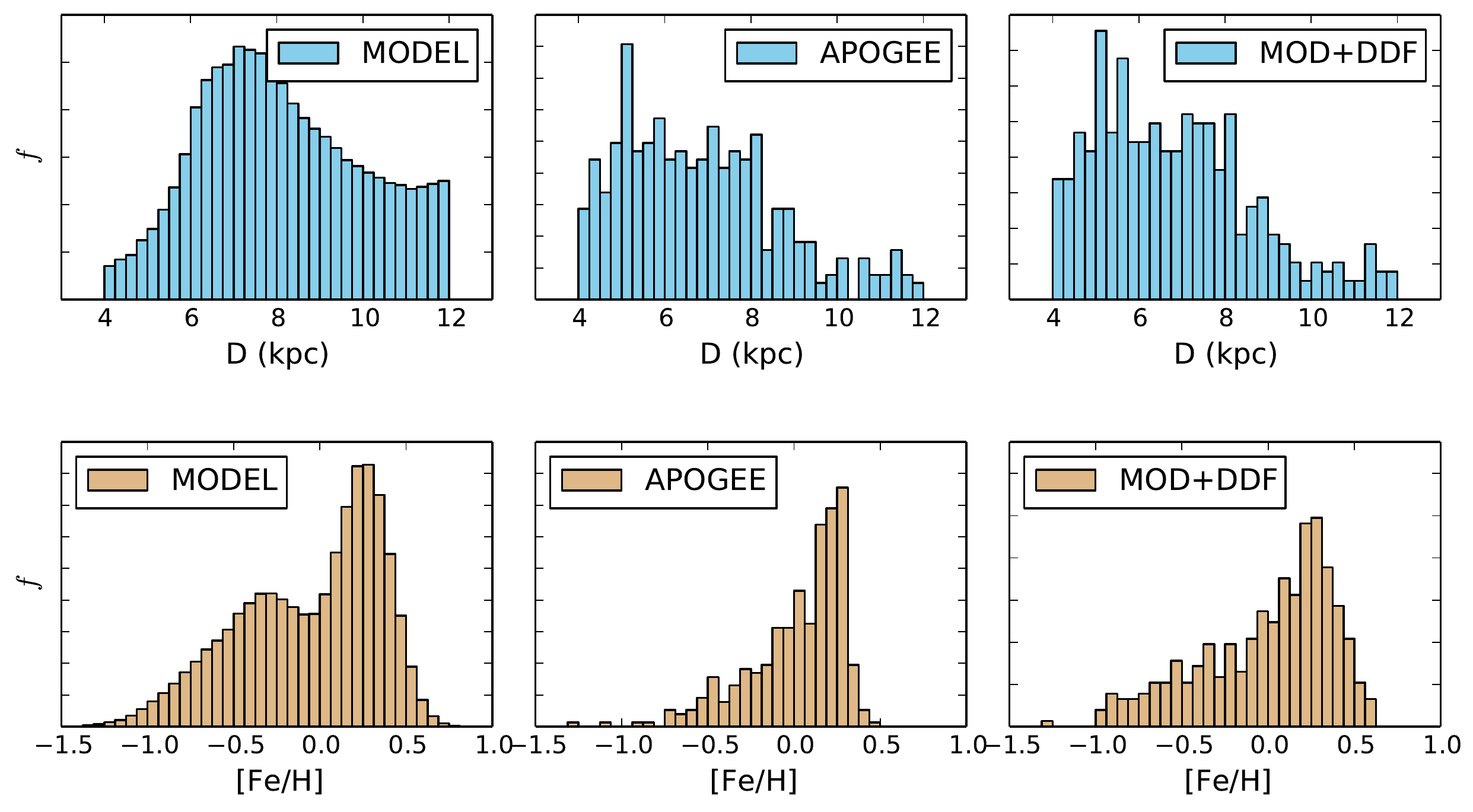}
\caption{Example of corrections applied to model in order to recover a sample with a distance distribution similar to APOGEE data in a given field. \emph{Top panels:} The distance distribution for the field centred around $l$ = 15, $b$ = 0, for the model (left panel), APOGEE DR13 (middle panel) and the model with the APOGEE distance distribution function (DDF) applied (right panel). \emph{Bottom panels:} The MDF from the model (left), from APOGEE DR13 (middle) and for the model with the APOGEE distance distribution function applied (right). We see that by applying the APOGEE distance distribution to the model we recover an MDF which better reproduces the data.}
\label{fig:modcor}
\end{figure*}

In this work we compare a number of the model predictions to data from the Apache Point Observatory Galactic Evolution Experiment (APOGEE; \citealt{Majewskietal2017}), a near-infrared, high resolution ($R\sim$22 500) spectroscopic survey of the Milky Way, which is able to probe the dust obscured bulge and inner disc regions. We make use of data release 13 (DR13; \citealt{SDSSDR13}) and use the distances derived in \cite{Wangetal2016}. We select APOGEE objects as recommended in the DR13 documentation (and see also Section 3 from \citealt{Wangetal2016} for the cuts they make to the data). For the metallicities we use calibrated global metallicities.

Directly comparing galactic models to data from spectroscopic surveys is not trivial, due to biases arising from the selection functions of the surveys. The selection function gives the fraction of stars observed in a given colour and magnitude range compared to the underlying stellar population of the Milky Way. It changes according to the targets selected (for example, giant or dwarf stars) as well as due to differences in the spectral resolution and wavelength coverage. In order to make a more faithful comparison of the model to the data we need to take the selection function into consideration; we do this by imposing the APOGEE distance distribution function (DDF), along different lines of sight, on the model. \cite{Nandakumaretal2017} showed that, for a given distance bin, different populations (e.g. giant or dwarf stars etc.) show similar trends in the MDF, i.e. that the resulting MDF is robust to the species of the tracer. This points to the fact that the distance distribution function is the main factor affecting the shape of the MDF. The particles in the simulation can be thought of as tracers of the underlying mass distribution, and therefore the selection fraction for each line of sight as a function of distance is the most important bias to take into consideration when comparing the model to the data.

An example of this type of ``forward modelling'' can be seen in Figure \ref{fig:modcor}, where we examine the distance distributions (top panels) and MDF (bottom panels) of a given region centred around $l$=15, $b$=0. In the left panels we show the distance distribution and MDF obtained by taking all the particles in a given field in the model; this is a perfect sampling of the underlying density distribution. In the middle panels we show the distance distribution and the MDF for the same field from the APOGEE data. We see that the distance distributions are very different for the model and the data, with the model distance distribution peaking at the centre of the galaxy while the data is biased to stars at the near side of the galactic centre, with the number of stars significantly decreasing at the far side. This kind of distance distribution will remove stars which are in the central kiloparsec of the galaxy compared to the model, as well as stars found preferentially at larger heights above the plane on the far side of the galaxy; in the case of our model, these are the metal-poor thick disc stars. This is reflected in the MDF's shown in the bottom panels; we see that the MDF from the model and observations are considerably different. The model predicts that there will be a significant contribution of the metal-poor component for the given field, while the APOGEE data does not have this metal-poor population. However, when we apply to the model the same distance distribution as in the APOGEE data\footnote{It's worth pointing out that when we apply the distance distribution of the APOGEE data to our model, we select the same number of particles in the model as the number of stars in the APOGEE data in the particular distance bin being considered. We found that for our purposes, forty distance bins were adequate for each field. The particles in each distance bin in the model are randomly selected, which introduces a certain degree of stochasticity when comparing the model+DDF to the data.} (top row, right panel), we see that the MDF of the model with the correction (bottom row, right panel) becomes much more similar to the MDF from APOGEE data. This is because a large chunk of metal-poor stars are removed from the field when this specific distance cut is applied.

In what follows, we will apply the distance distribution function to the model, when comparing it to the data. Apart from the ``pure'' model predictions, we will also show the model predictions with the APOGEE DR13 distance distribution function imposed to the model (dubbed MOD+DDF), to better understand which differences between the model and the data arise from biases in the observations and which arise due to differences between the underlying density distribution of the MW and the predictions of the model. 
In the MOD+DDF case we also apply additional ``observational'' errors in order to reproduce some of the errors in the data. Specifically, we add an error of 0.05 dex in metallicity and $\alpha$-abundances, and a 30\% dispersion on the distances in accordance to what is thought to be their error budget (see \citealt{Wangetal2016}).

%From Bovy+12: The observed density of G-type stars is simply the product of the underlying density with the sampling selection function, suggesting that one constrains this underlying density by forward modeling of the observations..

%----------------------------------------------------------------------------------------
%	SECTION density distribution
%----------------------------------------------------------------------------------------
\section{The Bulge and inner disc morphology} 
\label{sec:morph}

\begin{figure*}
\centering
\includegraphics[width=0.99\textwidth]{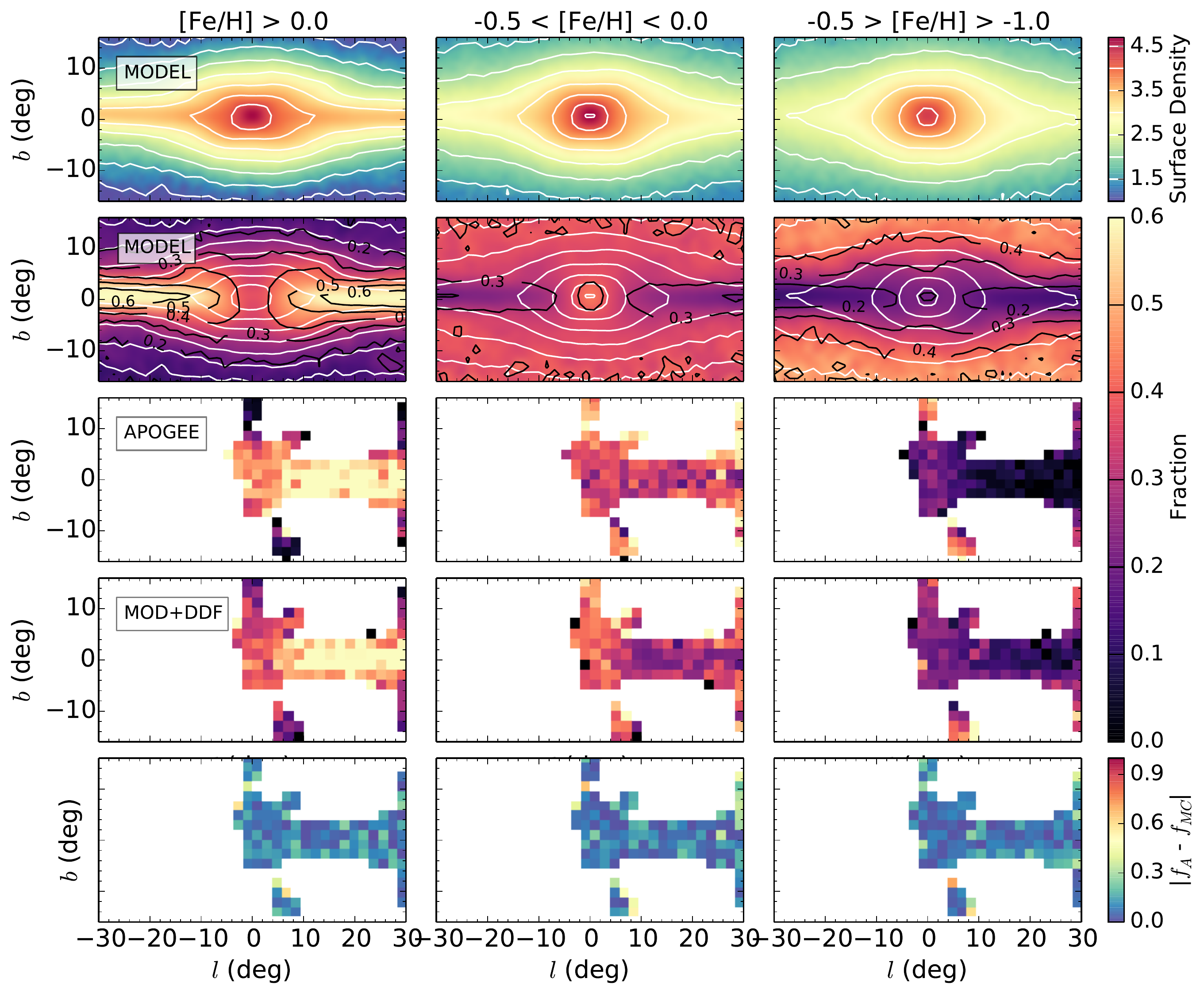}
\caption{\emph{Top row:} Surface density maps in $lb$ of the model after the formation of the bar and b/p bulge. The metallicity range of each population is indicated at the top (from left to right we show the metal-rich, intermediate and metal-poor component respectively). \emph{Second row:} Fractional contribution maps of each population in the model. \emph{Third row:} Fractional maps of the MW bulge and inner disc from the APOGEE DR13 data. \emph{Fourth row:} Fractional maps of the model+DDF for each component. \emph{Bottom row:} The difference in fractional contribution between the model+DDF and the APOGEE DR13 data.}
\label{fig:fraclb}
\end{figure*}

\begin{figure*}
\centering
\includegraphics[width=0.95\textwidth]{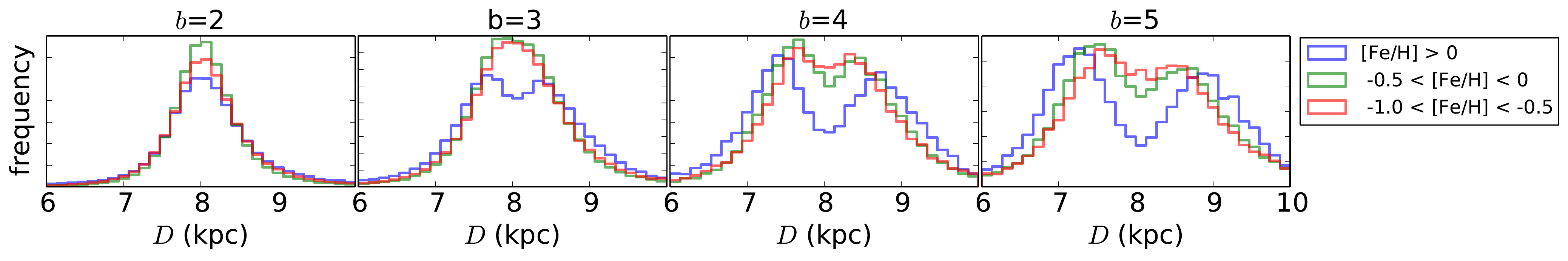}
\caption{The distribution of stars along the line-of-sight for four fields as indicated at the top of each panel along the galactic minor axis. Each curve corresponds to stars from the metal-rich (blue), intermediate (green) and metal-poor (red) disc.}
\label{fig:peanutlb}
\end{figure*}

\begin{figure}
\centering
\includegraphics[width=0.95\linewidth]{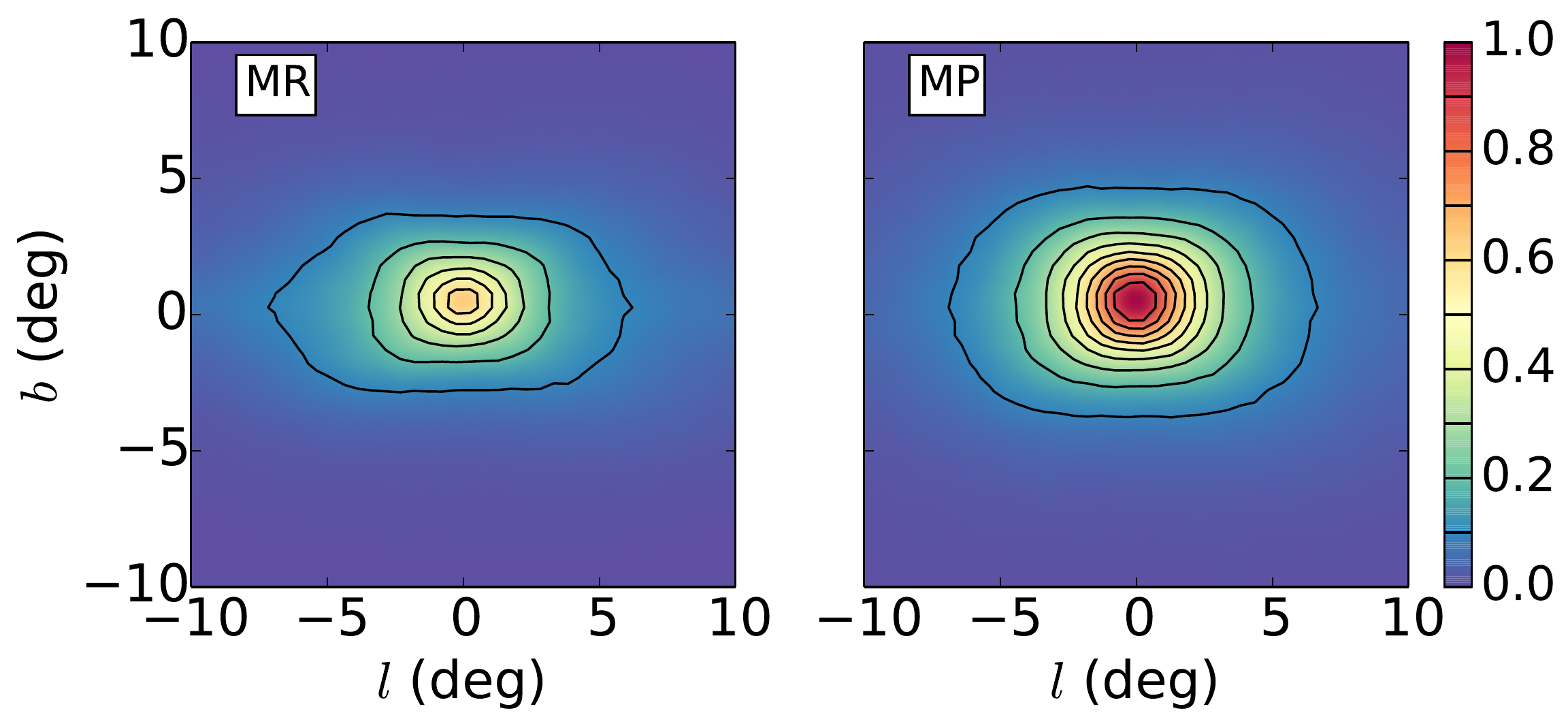}
\caption{Surface density plots in $l,b$, in linear scale for the metal-rich ([Fe/H] > 0; left) and metal-poor ([Fe/H] <0; right) components. The surface density plots are normalised by the maximum surface-density of the metal-poor component as in \citealt{Zoccalietal2017}.}
\label{fig:zoccali}
\end{figure}

\begin{figure*}
\centering
\includegraphics[width=0.98\linewidth]{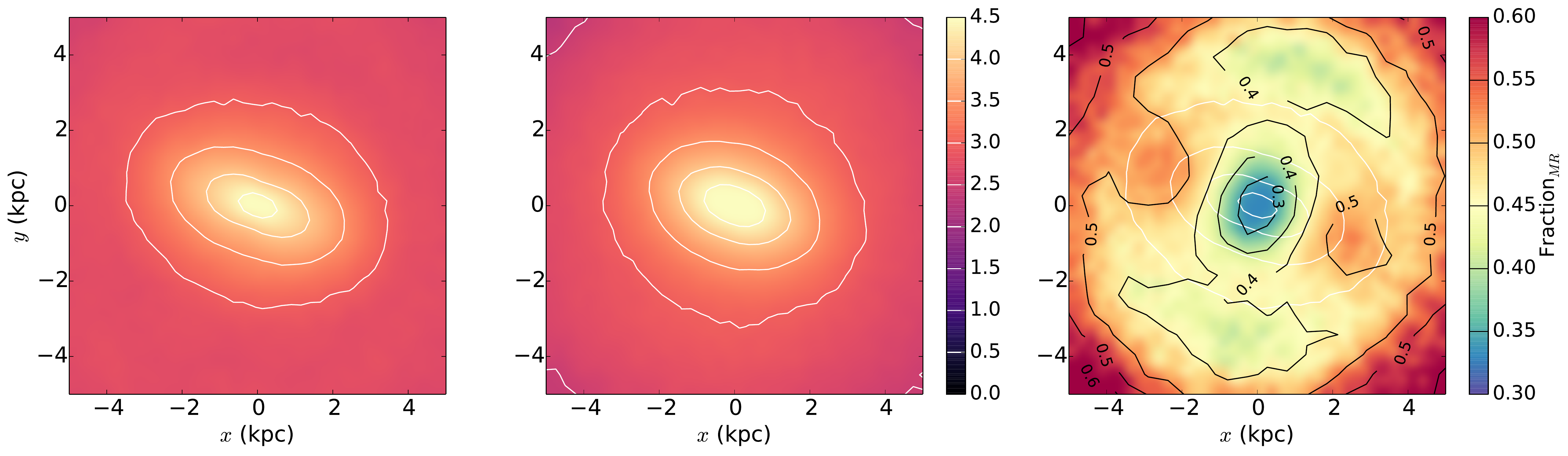}
\caption{Face on maps of the metal-rich component (left) and metal-poor components (middle panel). The right panel shows a fractional map of the contribution of the MR component. We see that there are azimuthal variations in the fractional contribution from the MR component in the dics.}
\label{fig:frac_faceon}
\end{figure*}

In this section we explore how secular evolution, i.e. how bar formation and the subsequent formation of the b/p bulge, redistribute the disc populations in our model, and how each of the populations is mapped into the b/p bulge and the inner disc.

%%%%%%%%%%%%%%%%
\subsection{Edge-on morphology}

In the bottom panel of Figure \ref{fig:model1} we show the global surface density of all the stellar particles of the model, as seen in the $l,b$ projection, with the bar at 30 degrees with respect to the galactocentric line of sight. We clearly see a prominent peanut-like shape inside |$l$| < 10 degrees, due to the presence of the b/p bulge. Our model is north-south symmetric since the Sun is placed in the plane of the galaxy. 
In Figure \ref{fig:fraclb} we separate the particles according to the disc population in which they originate, which corresponds to three different metallicity bins, as indicated at the top of each column. In the top panels we plot their surface density in $l,b$, where we see that the morphology of the b/p bulge is different in each of the populations and that the peanut shape is more prominent in the cold, metal-rich, thin disc population (as shown also in \citealt{DiMatteo2016,Athanassoulaetal2016,Debattistaetal2017,Fragkoudietal2017}). The hotter and more metal-poor populations on the other hand have rounder isophotes and appear less X-shaped. This occurs because particles originating in the hotter component are not trapped as efficiently by the bar as those originating in the cold component; the particles originating in the hot components thus lose less angular momentum than their colder counterparts and are trapped on less elongated orbits which have a weaker X-shape \citep{Fragkoudietal2017}. 

In the second row of Figure \ref{fig:fraclb} we show the fractional contribution of each of the discs to the total stellar mass. The white contours indicate the isodensity contours while the black contours indicate the fractional contribution (as indicated by the colourbar). We see that the three populations contribute differently at different heights above the plane. The cold disc population dominates close to the plane, with approximately 60\% contribution in the inner regions. The intermediate disc on the other hand has a more or less constant density fraction throughout the bulge region of about 30-40\% (see also \citealt{DiMatteo2016}), while the hottest population dominates at higher latitudes, i.e. above $\sim$10 degrees, where it contributes about 40\%. In the third row of Figure \ref{fig:fraclb} we show the fractional contribution of stars from APOGEE DR13, separating as in the model into the three metallicity bins indicated at the top of the figure. In the fourth row we show the MOD+DDF (see Section \ref{sec:compare}) to compare more directly with the APOGEE data. We see that the model shows very similar trends to the data; the metal-rich population ([Fe/H] > 0) dominates close to the plane (|$l$|>10), while the most metal poor population (-0.5 > [Fe/H] > -1) dominates further away from the plane. The intermediate population (0 > [Fe/H] > -0.5) is more or less constant in the bulge and inner disc region. We see that close to the plane in the bulge region, i.e. at |$l$|<10\,deg, the metal-poor populations are dominant over the metal-rich population. In the bottom panels of Figure \ref{fig:fraclb} we show the difference between the model and the data, in terms of fractional contribution; in most bins the difference is below 10\% with variations also due to noise and low number statistics.  

In Figure \ref{fig:peanutlb} we explore in more detail the signature of the peanut as seen in the three populations. We ``observe'' the bulge of the model along the minor axis, at different latitudes, as indicated at the top of each panel. We see that, at low latitudes, i.e. at $b$=2, the density distribution peaks at 8\,kpc, and there is no signature of the b/p in any of the populations, i.e. there is no double peak in the distribution of particles. As we move to higher latitudes, e.g. at $b$ = 3, the double peak of the b/p appears in the coldest component, while it is not evident in the hotter components. Therefore we see that as already evidenced in Figure \ref{fig:fraclb}, the peanut is more prominent in the metal-rich (cold) population, and therefore the signature of the peanut appears at lower latitudes in the thin disc than in the other components. As we move to even higher latitudes the double peak becomes evident also in the intermediate component, and at higher latitudes still, the double peak is eventually seen in all components. Additionally, we see that as we move up in latitude the separation between the peaks of the peanut increases (see also \citealt{Gomezetal2016}). In our model the peanut appears closer to the plane than the peanut of the MW bulge (see for example \citealt{Nessetal2013a}). This could be due to a number of reasons, such as the size and strength of the peanut not being the same as in the MW, which is an aspect of the N-body simulation that cannot be controlled in a straightforward manner.

In Figure \ref{fig:zoccali} we show the surface density of the metal-rich ([Fe/H] > 0) and metal-poor ([Fe/H] <0 ) components, normalised to the maximum density of the metal-poor population. We see similar features as those seen for the MW bulge with data from the GIRAFFE Inner Bulge Survey (GIBS) \citep{Zoccalietal2014}, as presented in Figure 9 of \cite{Zoccalietal2017}; firstly, the isodensity curves of the MP population are rounder than those of the MR population, which have a more flattened and boxy shape. The decreasing fraction of metal-poor populations close to the plane of the Galaxy is inverted at around $b$ = 5\,deg, and the metal-poor component becomes dominant again at low latitudes, i.e. the MP population has a higher surface density in the central region. In our model this is due to the fact that the intermediate and hot discs are centrally concentrated and thus dominate the mass budget in the innermost regions, accounting for over 50\% of the mass.

%%%%%%%%%%%%%%%%
\subsection{Face-on morphology}

In the left and middle panels of Figure \ref{fig:frac_faceon} we show the face-on morphology of the MR and MP populations. We see that the cold MR populations in the disc exhibit a stronger bar, i.e. are trapped more efficiently by the bar resonance than the hot populations, due to the different amount of angular momentum transferred from the cold and hot populations to the outer disc and halo (see \citealt{Fragkoudietal2017}). Therefore, the shape of the bar in the cold and hot discs is different, with the bar being more prominent in the cold populations, while it is rounder and weaker in the hot populations (and see also \citealt{BekkiTsujimoto2011,DiMatteo2016,Athanassoulaetal2017,Debattistaetal2017,Fragkoudietal2017}).

The fact that populations with different kinematic properties will respond differently to non-axisymmetric perturbations in the disc, will lead to azimuthal variations in the fractional contribution of the populations, wherever a non-axisymmetry is present (either in the bar or spiral arm region).
We see this in the right panel of Figure \ref{fig:frac_faceon}, where we show the fractional contribution of the MR population to the total stellar mass, for the inner few kpc of the model, as seen face-on. Due to the bar being more elongated in the MR population, the MR population dominates towards the ends of the bar along its semi-major axis, while outside the bar in the direction of the bar minor axis the MP population dominates. The MP population also dominates in the central-most region of the model, i.e. inside 1\,kpc, due to these populations being centrally concentrated and massive. Due to the different metallicities assigned in these components, the change in azimuth in fractional contribution naturally leads to azimuthal variations in metallicity in the inner few kpc of the model, as we show in Section \ref{sec:azimuth}.

%----------------------------------------------------------------------------------------
%	SECTION metallicity distribution
%----------------------------------------------------------------------------------------
\section{The Bulge and inner disc abundances} 
\label{sec:metal}

\begin{figure*}
\centering
\includegraphics[width=0.99\textwidth]{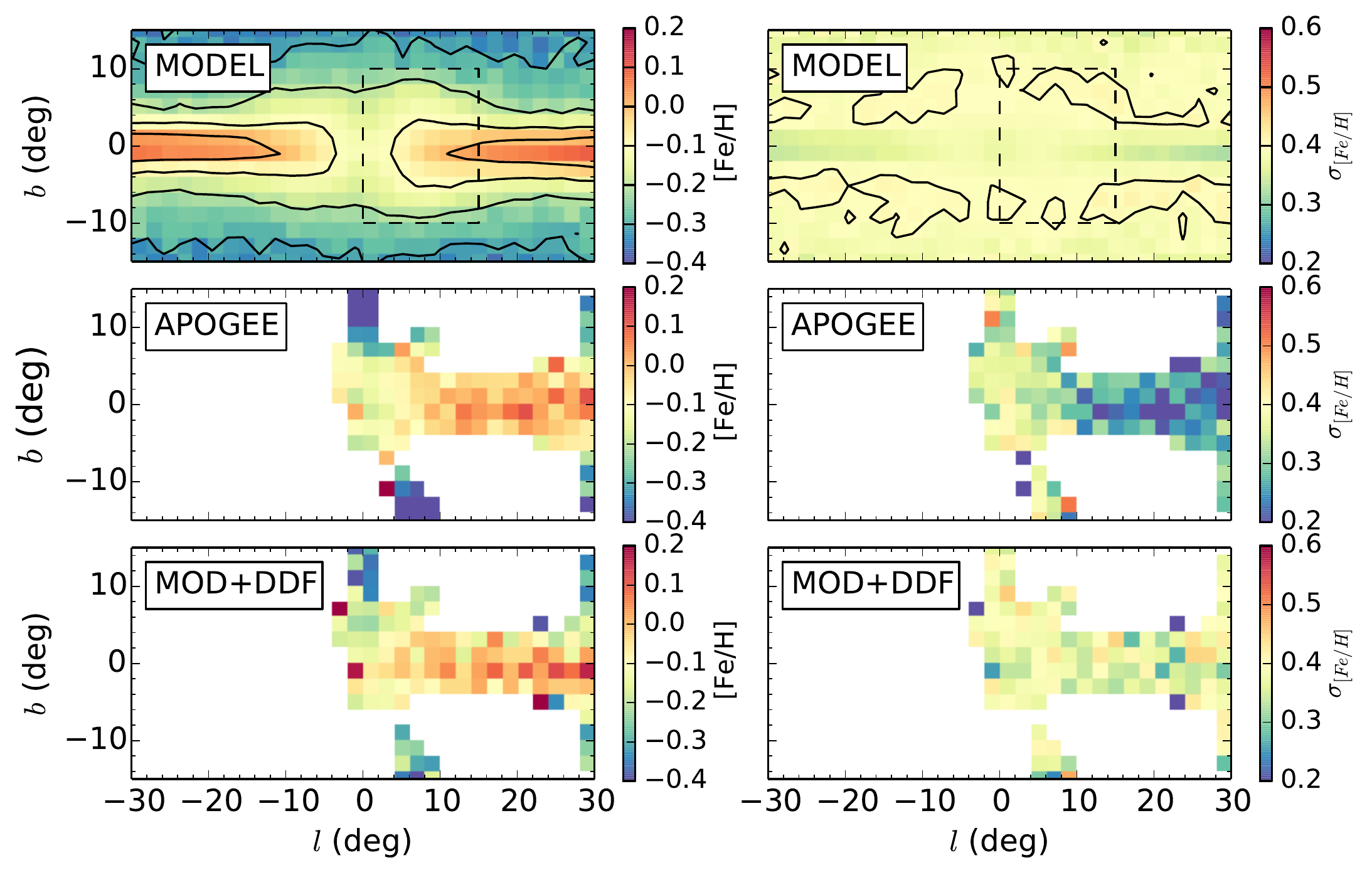}
\caption{\emph{Left panels:} Metallicity ([Fe/H]) along the line of sight for the model (top row), APOGEE DR13 (second row) and model+DDF (third row). \emph{Right panels:} Metallicity dispersion. For the APOGEE maps the number of stars used to construct the maps is 8585.}
\label{fig:meanfelos}
\end{figure*}

\begin{figure*}
\centering
\includegraphics[width=0.99\textwidth]{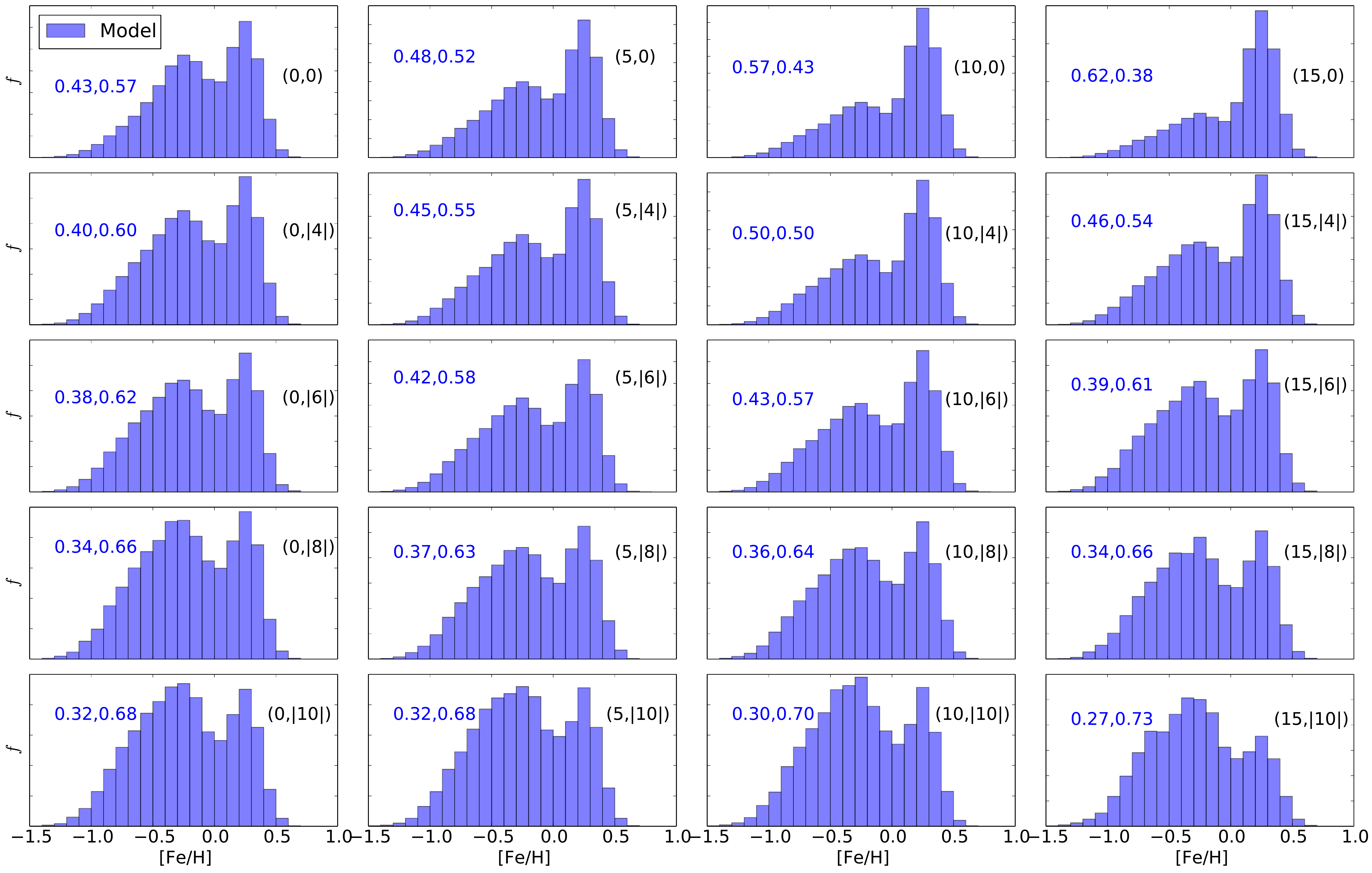}
\caption{MDF of the model (without the APOGEE distance distribution applied) for the fields outlined with a dashed line in Figure \ref{fig:meanfelos}, where the field is indicated in the top right corner of each panel. In the top left corner of the panels we show the fractional contribution of the MR ([Fe/H]>0) and MP ([Fe/H]<0) populations for each field examined. The histograms are normalised so that the total area under the curve is equal to 1.}
\label{fig:mdfbulge}
\end{figure*}

\begin{figure*}
\centering
\includegraphics[width=0.99\textwidth]{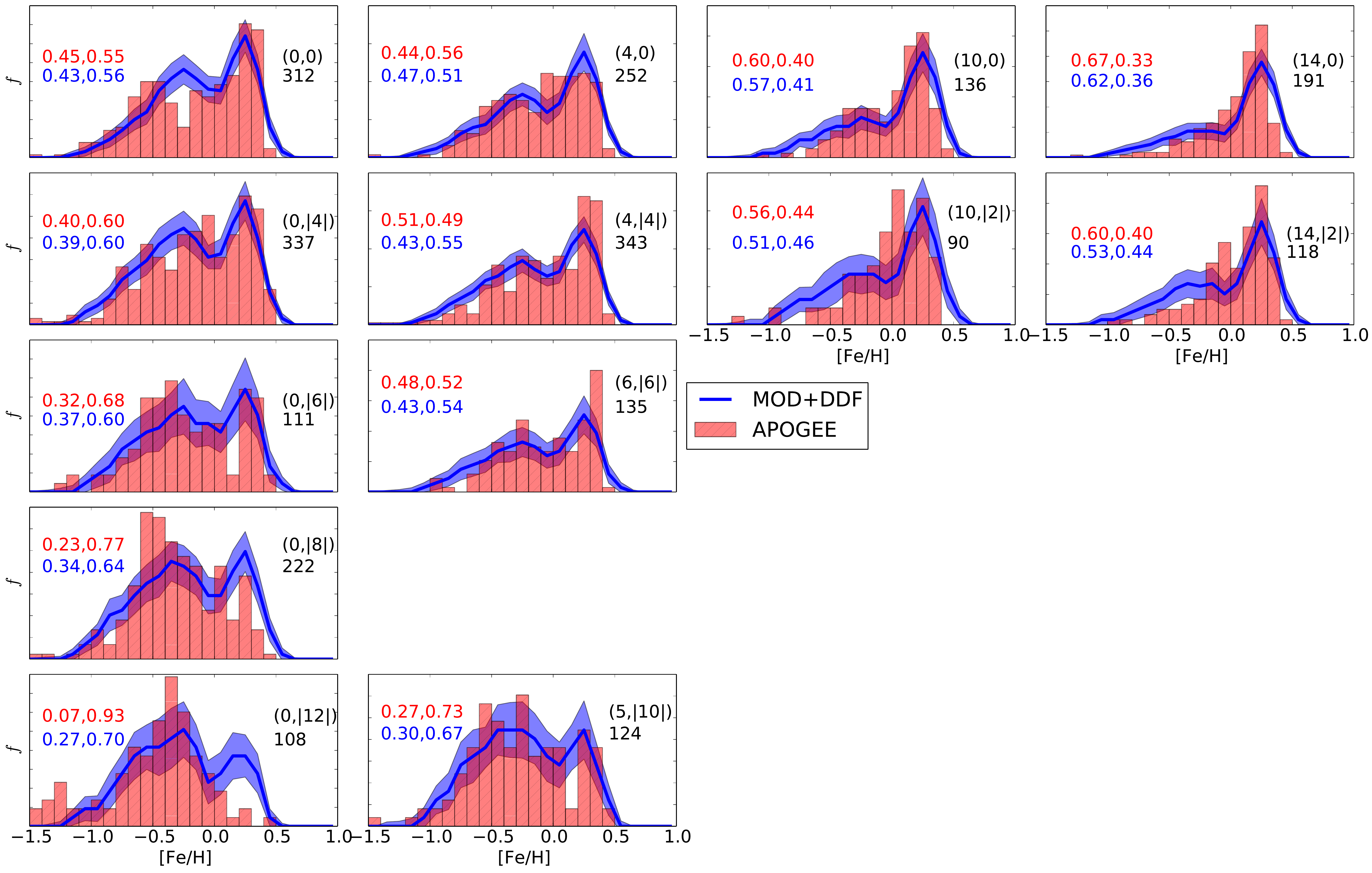}
\caption{MDF of the MW bulge using APOGEE DR13 data (red) compared to the MDF of the model+DDF (blue). For the model MDF+DDF we show the median (solid blue line) and dispersion (shaded blue area) for 100 realisations (see text for details). The field is indicated in the top right corner of each panel and below it we give the number of stars used in each field to construct the APOGEE MDF. In the top left corner of the panel we show the fractional contribution of the MR ([Fe/H]>0) and MP ([Fe/H]<0) populations for each field examined, for the APOGEE data (red) and for the model+DDF (blue).}
\label{fig:mdfbulgeDR13}
\end{figure*}

\begin{figure*}
\centering
\includegraphics[width=0.99\linewidth]{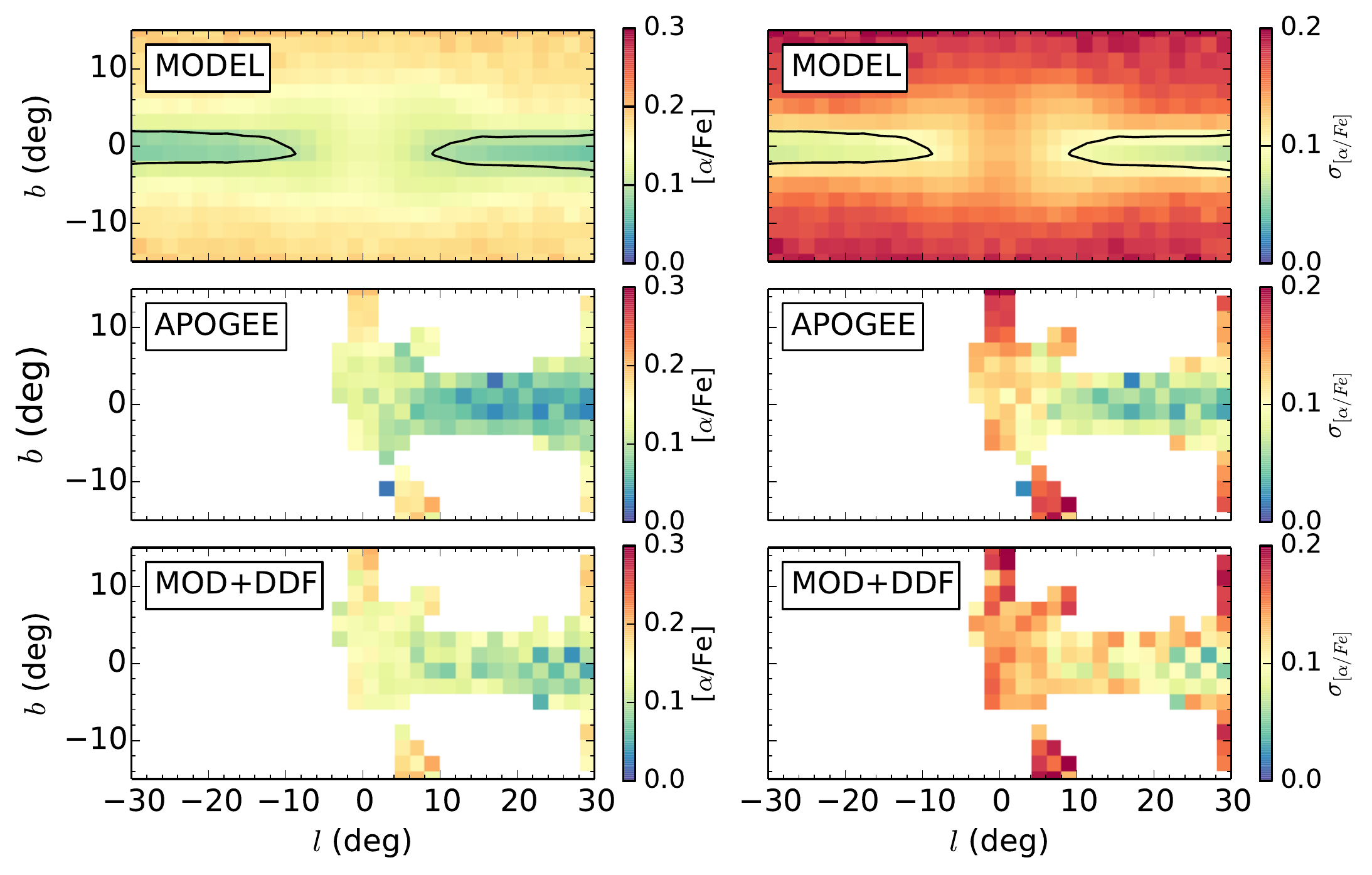}
\caption{\emph{Left panels:} Mean [$\alpha$/Fe] along the line-of-sight in $l,b$ for the model (top row), APOGEE DR13 (second row) and model+DDF (third row). \emph{Right panels:} [$\alpha$/Fe] dispersion. For the APOGEE maps the number of stars used to construct the maps is 8585.}
\label{fig:alphamean}
\end{figure*}

We explore the relation between morphology and abundances in the model, by examining maps of mean [Fe/H] and [$\alpha$/Fe], as well as the trends in the MDF, and compare these to data of the MW bulge from APOGEE DR13.

%----------------------------------------------------------------------------------------
%	SECTION average metallicity
%----------------------------------------------------------------------------------------
\subsection{Mean metallicity maps}

The first global photometric mean metallicity map of the MW bulge \citep{Gonzalezetal2013} revealed a mean metallicity of $\sim$ -0.1 dex at latitudes of $\sim$4 degrees, and confirmed the vertical negative metallicity gradient seen in spectroscopic surveys (e.g. \citealt{Richetal2007,Zoccalietal2008,Johnsonetal2011}). It also displayed the asymmetry seen in near-infrared images of the bulge \citep{Dweketal1995}, along with hints of a boxy shape. It did not however probe the regions close to the plane, due to crowding in the inner regions, thus leaving a gap in our knowledge of the metallicity of the innermost MW. This is now alleviated with the infrared spectroscopic survey APOGEE, with which we are able to probe regions close to the plane of the galaxy and construct metallicity maps at low latitudes.

In the left panels of Figure \ref{fig:meanfelos} we show the mean metallicity along the line-of-sight for the inner disc and bulge of our model (top panel), for APOGEE DR13 data (middle panel) and for our model+DDF (bottom row). In the right panels we show the metallicity dispersion for these three cases. In the top panels, the boxes drawn with dashed lines indicate the region in which we explore the bulge MDF in the next subsection.  

The APOGEE data show similar trends to what was seen in other spectroscopic surveys, i.e. there is a clear negative vertical metallicity gradient, while the bulge at latitudes of $b$$\sim$4 has a mean metallicity of the order of -0.1 dex \citep{Richetal2007,RojasArriagadaetal2014}. The APOGEE data also reveal that the bulge is metal-poor in the innermost regions (see also \citealt{Zoccalietal2017}), maintaining a mean metallicity of $\sim$-0.1, while it shows that the inner disc at $l$>10 degrees is metal rich, with the mean metallicity quickly rising to 0.1 dex. 

We see similar trends in our model, both in the top panel where we show the model without any observational errors or biases applied, as well as in the bottom panel where we apply the APOGEE distance distribution function. In our model the vertical metallicity gradient in the bulge arises due to the fact that the thick disc population dominates further away from the plane, while close to the plane in the innermost 10 degrees the bulge is metal-poor due to the combined contribution of the intermediate and hot discs (\emph{D2} and \emph{D3}). The thin metal-rich disc dominates close to the plane outside the bulge region, which is why the inner disc outside the bulge (i.e. $l>$10\,deg) is more metal-rich. 
We also see in the top left panel of Figure \ref{fig:meanfelos} that the metallicity map is more pinched than the underlying density distribution and that it has an X-shape (see also \citealt{Debattistaetal2017} and \citealt{Gonzalezetal2017} who discussed this in the context of the MW and external galaxies). 

In the right panels of Figure \ref{fig:meanfelos} we show the dispersion in metallicity. We see that the APOGEE DR13 data suggest that there is a metallicity dispersion of about 0.4 dex in the bulge region, which drops to values of 0.2-0.3 dex in the inner disc region, i.e. at $l$>10. Our model reproduces the metallicity dispersion in the inner region with values of about $\sim$0.4 dex. While there is a drop in metallicity dispersion in the model at $l$ > 10\,deg, we do not reach the low values of dispersion seen in the data. As we discuss in Section \ref{sec:limits}, this is possibly due to the discretisation of the continuous stellar populations of the MW disc into only three components. This means the most metal-rich particles (i.e. those of disc \emph{D1}) are distributed in a single, relatively thick, component (even though this disc is the thinnest disc in our model). 
%In our model we represent the whole thin disc with one single population with one single scaleheight and velocity dispersion, which of course is not the case in the MW, since the more metal-rich populations in the inner disc form over a period of 7-8\,Gyr when conditions in the interstellar medium change significantly. Therefore, to be more accurate we would likely have to have a part of the metal-rich population with smaller scaleheights and confined closer to the plane, thus contributing a larger number of metal-rich stars. In our case all metal-rich stars are found in a population with scaleheight a few hundred kpc.

%----------------------------------------------------------------------------------------
%	SECTION MDF
%----------------------------------------------------------------------------------------
\subsection{MDF of the Bulge}
\label{sec:metallicities}

In this subsection we explore the MDF for fields in the bulge and inner disc in the region indicated by the dashed box in Figure \ref{fig:meanfelos}. We compare the model and model+DDF predictions to the MDF obtained from APOGEE DR13 data.

In Figure \ref{fig:mdfbulge} we show the MDF of the model in various fields, with the field indicated in the top right corner of each panel. The size of the bins in this figure is 0.1\,dex and the size of the field is 2 and 4 degs in $b$ and $l$ respectively. To increase the number statistics we assume a north-south symmetry (which is reasonable for the model -- see Figure \ref{fig:model1}) and include particles from both positive and negative latitudes for a given field. In the top left corner of each panel we indicate the percentage of metal-rich ([Fe/H] > 0) and metal-poor ([Fe/H] < 0) stars in the given field.  
We see that there are variations in the MDF of the model bulge as a function of both longitude and latitude. Close to the plane, in the field $l,b$ = (0,0), the metal-rich population accounts for 40\% of the stellar mass, while the metal-poor population is dominant and accounts for $\sim$60\% of the stellar mass. In our model this is due to the concentrated metal-poor thick disc (\emph{D2} + \emph{D3}), and in particular due to the intermediate disc population which contributes $\sim$50\% of the surface density in these regions. Further out towards the inner disc of the model, i.e. at longitudes $l$>10, close to the plane (i.e. $b$=0), we see that the metal-rich population becomes more dominant, contributing approximately 60\% of the stellar mass. At higher latitudes, e.g. at $l$=0 and $b$=10, we see that the metal-poor population becomes more dominant, with 70\% of the mass budget attributed to it. As we move further out in longitude for large heights above the plane, e.g. at $l$=15 and $b$=10, the proportion of metal-poor stars increases still, reaching up to 75\% of the mass budget.
These changes in the MDF occur due to the changing weight with $l,b$ of the three different populations in our model, as indicated in the second row of Figure \ref{fig:fraclb}. 

In Figure \ref{fig:mdfbulgeDR13} we show the MDF of the MW bulge from APOGEE DR13 data in red, with the MDF from our model+DDF shown with the solid blue line. As explained in Section \ref{sec:compare}, the model+DDF is obtained by applying the APOGEE distance distribution function to the model along with some ``observational'' errors. This involves (stochastically) selecting particles according to the APOGEE distance distribution function in each field while the errors applied also induce some randomness in the produced MDF. We therefore calculate the MDF of the model+DDF by taking the median from 100 realisations (solid blue line) and also show the dispersion (shaded blue region) of these realisations. Most of the fields used have an area of 2\,deg$^2$. However the innermost field (0,0) is 4\,deg$^2$ and the rest of the in-plane fields (i.e. $b$=0) also have a width of 4 degrees in $l$ to increase the number of stars. For the fields of the inner disc, i.e., $l\geq$10, we split the $b$ = 2 APOGEE field to increase the statistics of the fields close to the plane. For example, in field (10,0) we include stars with |$b$| < 2 and in field (10,2) we include stars with 2 < |$b$| < 3.

The relative proportions of the MR and MP populations are given in the top left corner in red for the APOGEE DR13 data and in blue for the model+DDF. We see that the APOGEE data show variations in the MDF of the bulge as a function of both longitude and latitude, as has been shown also in other spectroscopic surveys of the MW bulge (e.g. \citealt{Nessetal2013a, RojasArriagadaetal2014, Zoccalietal2017}). The model+DDF shows very similar trends to those displayed by the APOGEE data. For the inner regions close to the plane (e.g. at $l,b$ =0), the metal-poor population contributes $\sim$60\% of the mass budget, while at higher latitudes its contribution increases further still. On the other hand, if we examine the inner disc region (e.g. at $l$>10), we see that the contribution from the metal-rich component increases and becomes dominant. 

The fractional contribution of metal-rich and metal-poor stars to the model MDF matches in most fields to within 10\% that of the data. In some fields, specifically at high latitudes (e.g. $l$=0, $b$=12) there can be a more significant mismatch. This is also related to the mismatch in metallicity dispersion between the model and the data. We discuss this in more detail in Section \ref{sec:limits}, but we point out here that this is likely due to the fact that the APOGEE MDF is more peaked towards high (low) metallicities at low (high) latitudes, compared to the model. 
%There also seems to be a mismatch in the central-most field, i.e. (0,0), where the model predicts a significant contribution form the intermediate disc, while the data seem to suggest that populations with 0 > [Fe/H] > -0.5 are not as significant at these latitudes. It is not clear why this mismatch occurs in the central-most degree, however we point out that by inspecting other close fields in the literature (e.g. field (0,-1) from the GIBS survey; \citealt{Zoccalietal2017}) there does not seem to be as much of a lack of intermediate metallicity populations. We therefore refrain from over interpreting this field at this point as future APOGEE data releases might release this issue.

%----------------------------------------------------------------------------------------
%	SECTION ABUNDANCES
%----------------------------------------------------------------------------------------
\subsection{The [$\alpha$/Fe] abundances}
\label{sec:abundances}

We explore the trends in [$\alpha$/Fe] abundances in the inner disc and bulge of our model and compare them to data from APOGEE DR13. We use the mean of [Mg/Fe] and [Si/Fe] to obtain an estimate of [$\alpha$/Fe] from APOGEE DR13.
In Figure \ref{fig:alphamean} the left panels shown the mean [$\alpha$/Fe] for the model (top row), for APOGEE DR13 data (second row) and for the model+DDF (bottom row). 

By examining the mean [$\alpha$/Fe] map obtained from the APOGEE data we see that in the bulge region the mean [$\alpha$/Fe] is of the order of $\sim$0.1 dex, with a positive vertical gradient, such that by latitudes of $b$$\sim$10 the [$\alpha$/Fe] reaches values of 0.2 dex. This trend is well-reproduced by the model, in which the vertical gradient arises due to populations with high [$\alpha$/Fe] (\emph{D3}) which dominate at larger heights above the plane. In our model the $\alpha$-poor populations are confined to the plane, where the thin disc is prominent, which leads to a low [$\alpha$/Fe] in the inner disc of the MW at $l$ > 10 degrees, of the order of 0.05\,dex. We also see that in the model the [$\alpha$/Fe] map has a pinched X-shaped distribution as is the case for the metallicity distribution. 

In the right panels of Figure \ref{fig:alphamean} we show the dispersion in [$\alpha$/Fe]. We see that the APOGEE data indicate that the [$\alpha$/Fe] dispersion in the bulge region is higher than in the inner disc, with values of $\sim$0.15\,dex in the bulge and $\sim$0.05\,dex for $l > $10. The model follows similar trends to the data, with a dispersion of $\sim$0.15 dex in the bulge, which decreases at higher longitudes. Similarly to the case of the metallicity dispersion, we see that while the overall match of the [$\alpha$/Fe] dispersion between the model and the data is very good, at latitudes $l$ >10 degrees the model dispersion is not as low as that seen in the data. As mentioned above, this is likely due to the discretisation of the thin disc in our model and we discuss this further in Section \ref{sec:limits}. However, we note that this is a second order effect and that most trends are very well reproduced with this simple model for the MW inner disc and bulge. 

%----------------------------------------------------------------------------------------
%	SECTION FACE-ON variations
%----------------------------------------------------------------------------------------
\section{Azimuthal metallicity variations in the inner disc}
\label{sec:azimuth}

\begin{figure*}
\centering
\subfigure[Face-on metallicity]{%
	\includegraphics[height=4.cm]{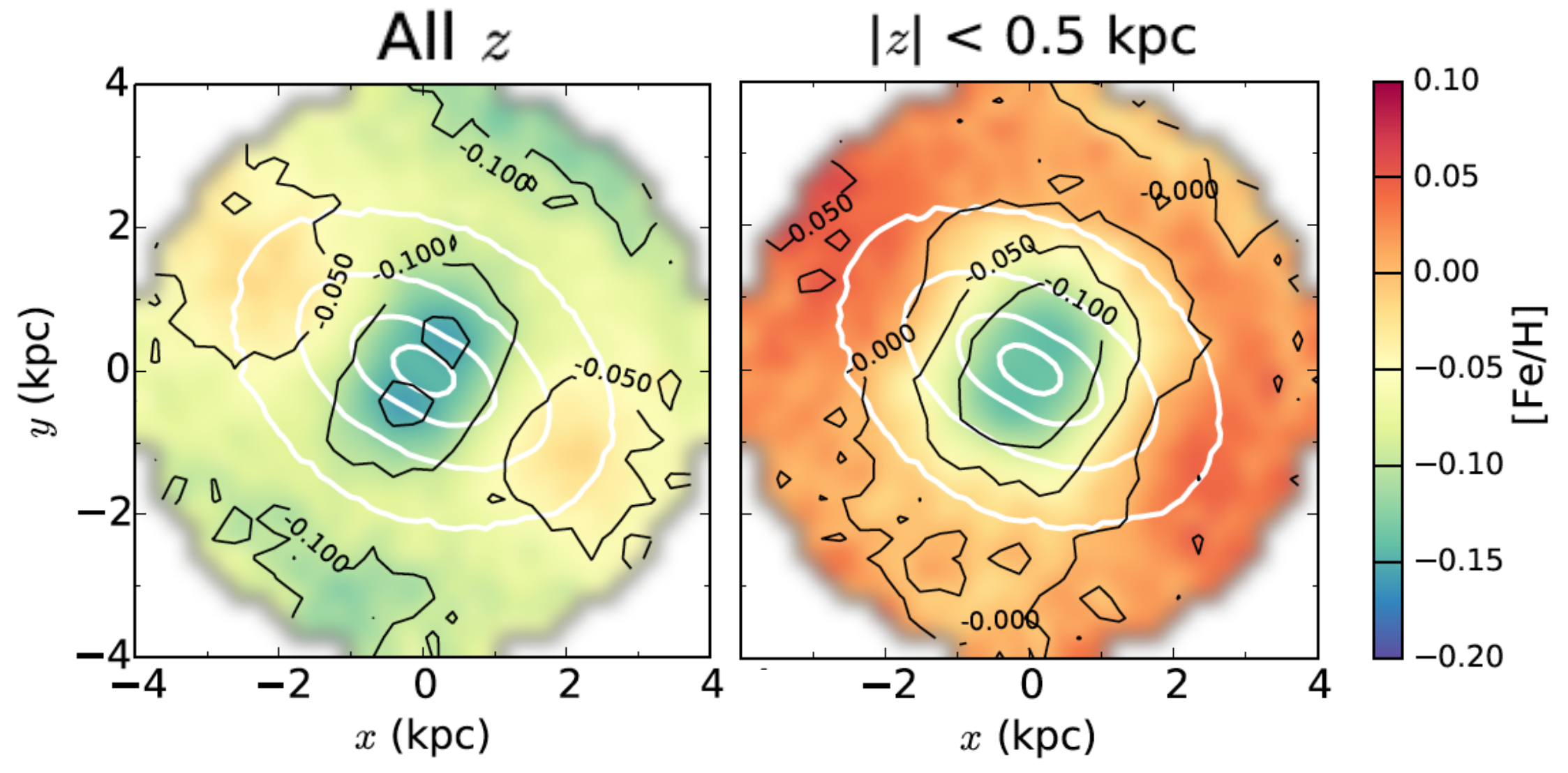}
	\label{fig:m1}}
\quad
\subfigure[Radial and azimuthal metallicity variations]{%
	\includegraphics[height=4.cm]{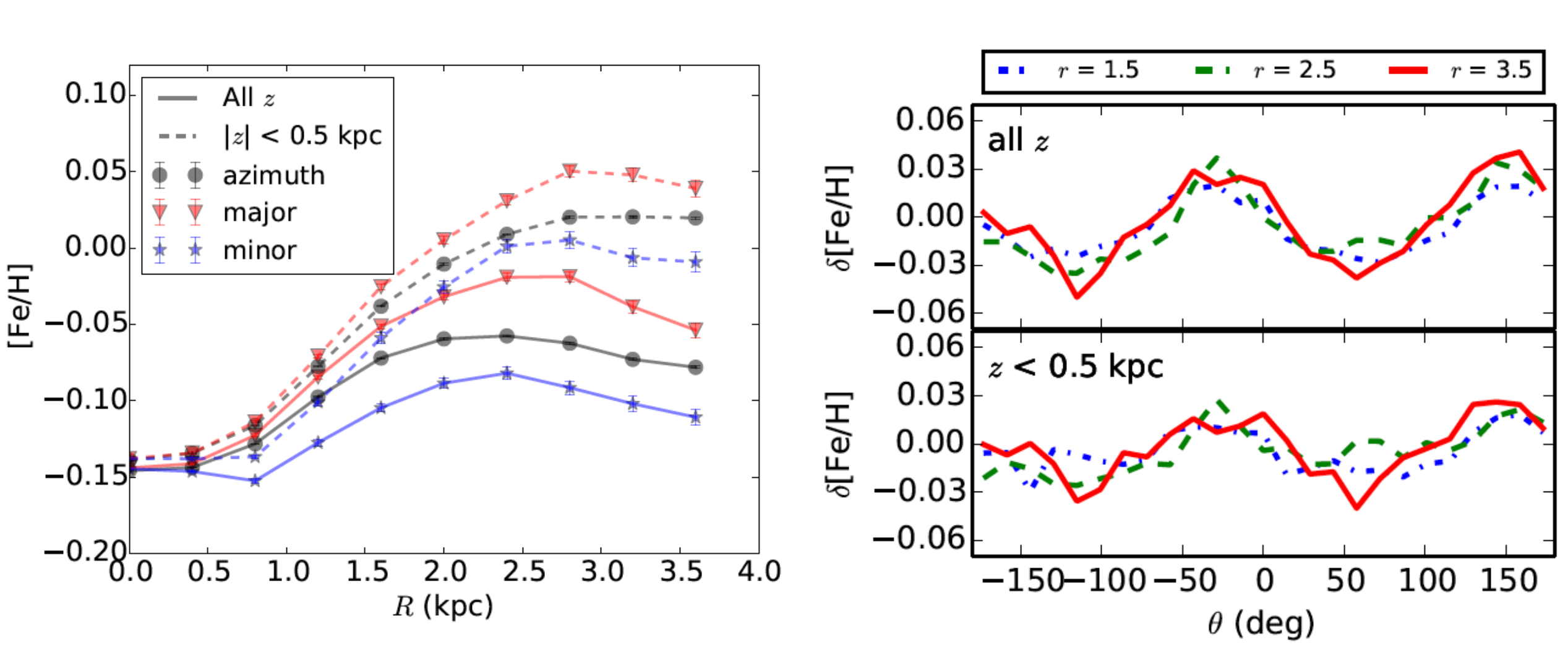}
	\label{fig:m3}}
\quad
\caption{\emph{a:} Face on metallicity map for all the particles in the simulation (left) and selecting only stars close to the plane, i.e. at |$z$| < 0.5\,kpc (right). \emph{b:} Metallicity as a function of radius, azimuthally averaged, along major and minor axis as indicated in the inset, for all stars and for stars close to the plane (left) and metallicity as a function of azimuth for different radii as indicated in the inset, for all particles in the simulation and for those within |$z$| < 0.5\,kpc from the plane.} 
\label{fig:FeH_faceon}
\end{figure*}

In this section we explore the azimuthal metallicity variations of the inner disc and bulge of our model when observing it face-on. Something interesting to note is that the mechanism for these variations arises from how cold (metal-rich) and hot (metal-poor) stellar populations respond to non-axisymmetric instabilities in the disc, in this case the stellar bar. 

In Figure \ref{fig:FeH_faceon} we show the azimuthal metallicity variations, and the overall metallicity distribution in face-on projections of the model. In the left panel of Figure \ref{fig:m1} we show the average metallicity of the model, obtained by taking all stars along $z$, when observing the model along the $z$ axis, i.e. face-on. The isodensity contours of the bar are shown in white, while the iso-metallicity contours are shown in black. We see that there are indeed metallicity variations of up to 0.1\,dex in the inner disc; the innermost regions are metal-poor, due to the high fractional contribution of the massive metal-poor intermediate and hot discs, which have short scalelengths and are thus centrally concentrated. This can be understood also by examining the map of the face-on fractional contribution of the metal-rich population in Figure \ref{fig:frac_faceon}, where we see that in the innermost kiloparsec the MR population is subdominant. On the other hand, at the edges of the bar, where there is a higher fraction of the thin, metal-rich population, the global metallicity increases. Along the bar minor axis the metallicity is relatively low compared to the edges of the bar major axis, since the hot populations dominate there. 
In the right panel of Figure \ref{fig:m1} we select stars which are close to the plane, i.e. $|z|$<0.5\,kpc. We see that there are still azimuthal variations, although they are not as pronounced as in the left panel of Figure \ref{fig:m1}. The disc outside the central kiloparsec is also on average more metal-rich, since we are preferentially selecting stars close to the plane i.e. from the thin metal-rich disc component. 

In the left panel of Figure \ref{fig:m3} we show the mean metallicity as a function of radius, along the major and minor axis of the bar and azimuthally averaged, with red, blue and black symbols respectively. We see that there are clear variations in the metallicity gradient depending on if we measure the gradient along the bar major or minor axis. In the right panel of Figure \ref{fig:m3} we show the azimuthal metallicity variations for different radii in the inner disc. In the top panels of the plot we select all stars in $z$ while in the bottom panel we select stars with |$z$|<0.5\,kpc. The azimuthal metallicity variations are calculated by comparing the metallicity in an azimuth bin to the mean azimuthally averaged metallicity, i.e. $\delta$[Fe/H] = [Fe/H]$_{\theta}$ - <[Fe/H]>. We see that there are variations in azimuthal metallicity of up to 0.1\,dex, which could be measurable with current and upcoming spectroscopic surveys. 

We emphasise that the aforementioned radial and azimuthal metallicity variations in the inner disc (when viewing the model face-on) are likely due to how the cold and hot components respond to the bar perturbation. The different kinematics of the populations lead to a bar which is more prominent and elongated in the cold component compared to the bar in the hot component (see also \citealt{Fragkoudietal2017} for a discussion on the morphology of bars in composite cold/hot discs and Khoperskov et al. in prep. for a discussion of this effect on azimuthal variations in spiral arms).

%----------------------------------------------------------------------------------------
%	SECTION discussion
%----------------------------------------------------------------------------------------

\section{Discussion}
\label{sec:discussion}

%----------------------------------------------------------------------------------------
%	SECTION vertical metallicity gradient
%----------------------------------------------------------------------------------------
\subsection{The origin of the vertical abundance gradients}
\label{sec:gradeint}

\begin{figure}
\centering
\includegraphics[width=0.9\linewidth]{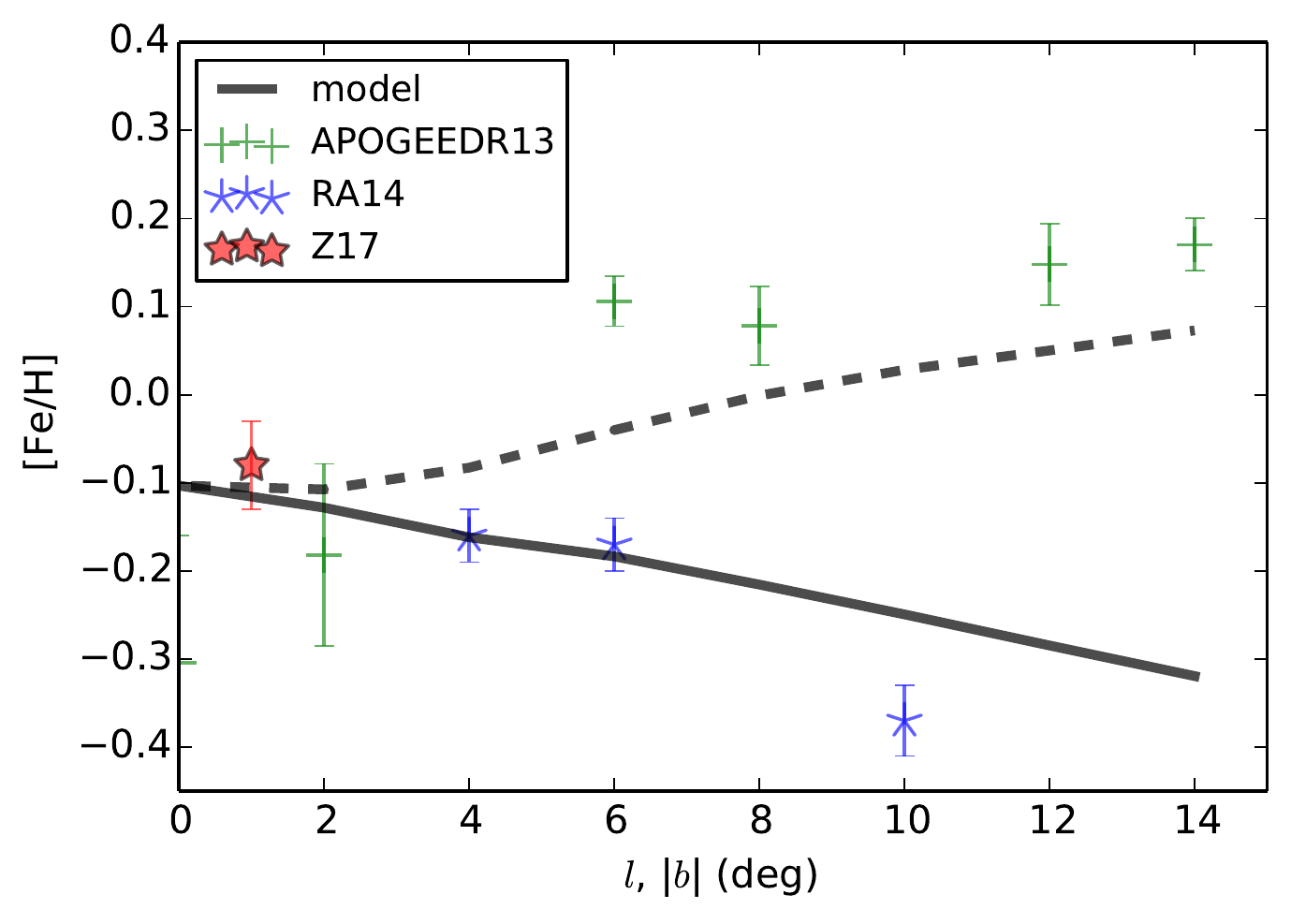}
\caption{Vertical (solid) and longitudinal (dashed) metallicity gradients of the bulge in the model, compared to mean metallicities of fields along the minor axis and along the $b$=0 axis, from spectroscopic surveys APOGEE DR13, Gaia-ESO survey (GES, \citealt{RojasArriagadaetal2014}) and the Giraffe Inner Bulge Survey (GIBS, \citealt{Zoccalietal2017}).}
\label{fig:gradlb}
\end{figure}

We show in Figure \ref{fig:gradlb} the vertical metallicity gradient (solid line) of the bulge in the model, as well as the longitudinal (dashed line) gradient of the bulge. We compare to data from APOGEE DR13, which was used in the rest of this study, as well as to data from the Gaia-ESO \citep{RojasArriagadaetal2014} and GIBS survey \citep{Zoccalietal2017}\footnote{We use the mean and error on the mean for relevant fields along or close to the minor axis of the bulge from the articles cited. Specifically from \citealt{Zoccalietal2017} we use field (0,-1), while from \citealt{RojasArriagadaetal2014} we use fields (1,-4), (0,-6) and (-1,-10)}. We see that there is a clear positive longitudinal gradient in the MW and in the model, as well as a negative vertical gradient in the MW bulge. 

In our model the vertical gradient has a value of -0.015\,dex/deg. This fits the data well for the inner 6 degrees, however the data show a steeper gradient between $b$ = 6-10\,deg. The longitudinal gradient predicted by the model is positive, with a value of 0.014\,dex/deg. The slope of the model seems to match the slope of the data, however the data show a zero-point offset to slightly higher metallicities. Some of the reasons for this mismatch are discussed below in Section \ref{sec:limits}.

There have been various mechanisms proposed to explain the origin of the vertical metallicity gradient of the MW bulge. These include invoking a classical bulge, as it can naturally explain the radial gradient in metallicity \citep{Zoccalietal2008}. Other studies have shown that the redistribution of stars by a bar and b/p bulge, can also produce this vertical metallicity gradient as long as the disc has a steep initial radial metallicity gradient (see for example \citealt{BekkiTsujimoto2011,MartinezValpuestaGerhard2013,DiMatteoetal2014}). However, as we showed in recent work \citep{Fragkoudietal2017b}, while such models can indeed reproduce the vertical metallicity gradient, they do not reproduce a number of other trends seen in the MW bulge and inner disc, such as the metal-poor innermost regions and the positive longitudinal gradient. The positive longitudinal gradient seen in the inner MW arises naturally in our model due to the metal-rich thin disc which becomes dominant at longitudes $l>$10.

In our model, the vertical metallicity gradient is present in the initial conditions of the disc and is due to the different scaleheights of the various disc populations, from thin and metal-rich, to thick and metal-poor. The metal-rich disc thus dominates close to the plane, and the metal-poor disc dominates at larger heights as we showed in previous sections (and see also for example \citealt{BekkiTsujimoto2011}). These populations are redistributed via the formation of the bar and the b/p, and the steepness of the metallicity gradient is changed by the formation of the b/p bulge. The vertical metallicity gradient in our model is therefore due to the presence of the thin and thick disc in the inner parts of the galaxy. We refer the reader to \cite{Fragkoudietal2017b} for a more detailed comparison between this model and a model with an initial steep radial metallicity gradient. 

Furthermore, we also see in Figure \ref{fig:alphamean} that the MW bulge has a vertical abundance gradient. This gradient arises in a similar fashion to the metallicity gradient (although inverted) since the metal-poor thick disc populations are also $\alpha$ abundant, while the metal-rich thin disc populations are $\alpha$-poor. It is worth pointing out that, since the stars in the MW bulge which are metal-poor are also $\alpha$-abundant, in a scenario in which the vertical-metallicity gradient of the MW bulge were due to an initial radial metallicity gradient in the disc, this would imply a steep positive $\alpha$-abundance gradient in the inner disc.

%By these two arguments, we see that it is unlikely that the vertical metallicity gradient in the Milky Way bulge was due to a steep metallicity gradient (-0.4 dex/kpc) in the disc at high redshift.

%----------------------------------------------------------------------------------------
%	SECTION metal-poor inner regions
%----------------------------------------------------------------------------------------
\subsection{The metal-poor innermost region}

We see by examining the mean metallicity maps in Figure \ref{fig:meanfelos} as well as the MDF of the inner regions of the model bulge (Figure \ref{fig:mdfbulge}) that the inner few degrees, between |$l,b$| < 4, are dominated by the metal-poor component. In the inner regions the metal-poor component contributes more than 50\% of the mass (see the fractions given in Figure \ref{fig:mdfbulge}). The majority of this contribution in the MP component arises due to contributions from the intermediate (\emph{D2}) and hot (\emph{D3}) disc populations, which contribute $\sim$50\% and 10\% of the mass in the innermost region respectively. Therefore the MP component in total contributes approximately 60\% to the total mass in the innermost few degrees. This leads to a mean metallicity of the order of -0.1\,dex, consistent with what is seen in the APOGEE data.

It is important to emphasise that with this fairly simple model, we are able to reproduce the metal-poor innermost regions seen in the MW bulge. This shows that it is possible to explain the majority of the stellar populations in the bulge in terms of disc populations, without the need to add neither a classical bulge component, nor a stellar halo component (which would likely contribute only few percent to the mass budget). This arises in the model because the scalelength of the intermediate and hot discs are short, and importantly, because the discs are massive, representing 50\% of the stellar mass of the Milky Way.

%----------------------------------------------------------------------------------------
%	SECTION vertical metallicity gradient
%----------------------------------------------------------------------------------------
\subsection{The number of peaks in the MDF}
\label{sec:peaks}

A discussion which is undertaken often in the literature regards the number of gaussians, or components into which the MDF of the bulge can be separated. For example, \cite{Schultheisetal2017} find an MDF with only two components in Baade's window, as is also found for example by the APOGEE and Gaia-ESO surveys \citep{GarciaPerrezetal2017,RojasArriagadaetal2014}, while in other surveys, for example the ARGOS survey \citep{Freemanetal2013} the authors find three peaks in the MDF \citep{Nessetal2013a}. This has stirred up debate regarding how many separate, physically motivated components, there are in the inner regions of the MW.

Our model has three disc components -- which are a coarse discretisation of the continuous mono-abundance populations seen in the MW disc --, and each has a metallicity assigned to it in terms of a gaussian with a given mean and dispersion. Even though we have three components in our model, with distinct morphological, kinematical and chemical properties, the final MDF only has two peaks, as can be clearly seen in Figure 2. We see therefore, that the number of peaks will depend on the choice of dispersion in metallicity for each component, irrespective of the number of components; had we used lower dispersion for our intermediate and hot discs, we would have recovered an MDF with three rather than two peaks. Having an $n$ number of peaks in the MDF therefore does not necessarily mean that there are $N$ physically different components in the galaxy; we could have added any number of $N$ discs to our model, and ended up with any number of $n$ components, depending on the mean and dispersion we assign to each component. Therefore, we see that the number of peaks $n$ does not necessarily point to physically distinct components.

%----------------------------------------------------------------------------------------
%	SECTION the radial-longitudinal metallicity gradient
%----------------------------------------------------------------------------------------
%\subsection{The longitudinal metallicity gradient}

%----------------------------------------------------------------------------------------
%	SECTION azimuthal metallicity variations
%----------------------------------------------------------------------------------------
%\subsection{Azimuthal metallicity variations}
%\label{sec:az}

%----------------------------------------------------------------------------------------
%	SECTION drawbacks
%----------------------------------------------------------------------------------------
\subsection{Limitations of the model}
\label{sec:limits}

The model presented in this work is a simple model with three disc components, which aims to describe a large fraction of stellar populations in the inner disc and bulge of the MW. The three discs presented here are a first step to discretising the continuum of stellar populations seen in the disc of the MW \citep{Bovyetal2012}. The three discs do not correspond to three \emph{distinct} components of the MW, however, they can roughly be assigned to different phases in its star formation history (see \citealt{Snaithetal2014}) and correspond to the metal-rich thin disc, the young thick disc and the old metal-poor thick disc -- nomenclature as in \cite{Haywoodetal2013}. 

While the model is quite successful in describing how secular evolution can redistribute stars from the thin and thick disc into the b/p bulge, it does not include continuous star formation and stellar feedback processes, nor any gaseous physics. Discretising a continuum of stellar populations into three disc populations is of course a rough approximation. In fact our cold/thin disc component (\emph{D1}) is representative of the last 8\,Gyrs of evolution of the MW disc, when of course not all stellar populations formed in this time will have the same kinematics and morphologies, since populations form with cooler kinematics at lower redshifts (e.g. \citealt{Birdetal2013,Wisnioskietal2015}). Therefore, there are likely populations in the MW which are formed in the last 8\,Gyrs which are kinematically colder and thinner than the single disc population used to characterise them in this model. 

This coarse discretisation of the continuous populations of the disc becomes most evident when examining fields with  $l\geq$10 at low latitudes. There, the data show a MDF which is peaked towards high metallicities, while the model has a longer tail of metal-poor populations. This is likely not due to an over-abundance of metal-poor stars, but due to the fact that there should be a higher relative fraction of metal-rich stars close to the plane in the inner disc. This also produces a mismatch, as we saw in previous subsections, between our model metallicity dispersion and that of the observations. If there were ongoing star formation in the model, then there would be populations which at late times would contribute to the populations with higher metallicities close to the plane, which would make the MDF there more peaked at higher metallicities. This would subsequently also lead to a smaller metallicity dispersion in the inner disc of the model.

There is also a slight mismatch between the APOGEE and modelled MDF at high latitudes, i.e. at $b =$12 where we see the inverse problem, i.e. an overabundance of metal-rich stars compared to metal-poor stars. This again is likely due to the coarse modelisation of 8\,Gyrs of evolution of disc populations in a single component. The mismatch occurs again because there is likely a part of the metal-rich cold thin disc populations which, instead of being confined close to the plane -- which would happen naturally with later star formation, are transported to larger heights by the bar-b/p instability and thus contribute also at high latitudes. 

While we will present models with continuous star formation in future papers, it is worth emphasising that in general this simple model matches the overall trends of the MW bulge and inner disc surprisingly well, and that the aforementioned points are secondary in importance.

%This is likely due to the fact that we use one single disc population to represent the entire thin disc (since there is no gas and star formation in our model). If we introduced star formation in the model this would contribute to thinner and more metal-rich populations close to the plane. By changing the proportion of stars to higher or lower metallicities would contribute to a more peaked MDF at the metal-rich and metal-poor end respectively, which would lead also to a lower metallicity dispersion. 

%The lack of star formation is also likely the reason for the over abundance of metal-rich stars further away from the plane, which can be seen in the bottom fields in Figure \ref{fig:mdfbulgeDR13}. Indeed it is likely that due to the fact that we characterise all the thin disc (i.e. populations born in the last 8 Gyrs) with one single population, which all buckle when the bar forms and buckles

%----------------------------------------------------------------------------------------
%	SECTION summary
%----------------------------------------------------------------------------------------

\section{Summary \& Conclusions}
\label{sec:summary}

In this paper -- part of a series exploring the connection between the disc and bulge of the Milky Way (MW) -- we examine the chemo-morphological relations of stellar populations in the MW bulge. 
To this aim, we study an N-body simulation of a composite disc MW-type galaxy -- with both a thin and a massive and centrally concentrated thick disc -- which evolves in isolation, subsequently forming a bar and a boxy/peanut (b/p) bulge. We compare the predictions of the model to data from the near-infrared APOGEE survey. 

The composite disc of the model is made up of three discs, which represent a discretisation of the continuous stellar populations seen in the disc of the MW \citep{Bovyetal2012}. These have morphologies, kinematics and metallicities characteristic of the metal-rich thin disc and the metal-poor thick disc of the inner MW (see \citealt{DiMatteo2016} for more details on this scenario). The bulge in this model is therefore made up of \emph{inner} disc stellar populations, which are mapped into the b/p through secular processes, i.e. the formation and vertical heating of a stellar bar.

We show (in Figure \ref{fig:modcor}) the importance of taking the selection function of the survey into account when comparing the model with the data, specifically the distance distribution function. We ``degrade'' the simulation by taking into account the distance distribution of the APOGEE survey in the fields examined, in order to make a more accurate comparison of the model to the data. 

We construct metallicity and [$\alpha$/Fe] maps, and examine the MDF in different fields of the bulge in our model, and then compare to those extracted from APOGEE data. We show that the model is able to reproduce a number of observables in the MW bulge and inner disc. These include:
\begin{itemize}
\item The rounder shape of the metal-poor (MP; [Fe/H] < 0) populations compared to the metal-rich (MR; [Fe/H] > 0) populations in the bulge
\item The inversion in the fraction of MP/MR stars at low latitudes, as observed in the GIBS survey \citep{Zoccalietal2017}, leading to a higher fraction of MP stars in the centralmost regions of the bulge
\item The overall trends in the mean [Fe/H] and [$\alpha$/Fe] maps derived from APOGEE DR13 data
\item The trends in the MDF of the bulge and inner disc, as a function of longitude and latitude
\item The vertical metallicity gradient, the positive longitudinal metallicity gradient and the metal-poor inner regions of the bulge (see also \citealt{Fragkoudietal2017b})
\item The model also predicts that there are small but measurable azimuthal metallicity variations in the inner disc, of the order of 0.1\,dex, as well as differences in the radial gradients measured along the bar major and minor axes. These arise due to the differential mapping of cold and hot populations in the bar-b/p 
\end{itemize}

In our model, all the aforementioned are due to the mapping of a thin and thick disc into the b/p, and the final morphology and densities of these disc populations at different longitudes and latitudes (see Figure \ref{fig:fraclb}). To reproduce the trends, the metal-poor and $\alpha$-enhanced populations must be massive and centrally concentrated; this has in fact been shown to be the case for the MW (for the scalelength see for example \citealt{Bensbyetal2011,Bovyetal2012} and for the mass see \citealt{Snaithetal2014,Haywoodetal2015}). These morphologies occur due to the differential mapping of cold/thin and hot/thick disc populations in the b/p bulge (see \citealt{Fragkoudietal2017} for a discussion on the physical processes driving this differential mapping).

In this model most of the stellar populations in the bulge (with [Fe/H] $>$ -1) are naturally accounted for in terms of the cold, metal-rich thin disc and the hot, metal-poor, thick disc populations seen in the inner Milky Way. While some of the trends explored in this work can also be explained by a massive classical bulge, a number of works have shown that such a massive spheroid cannot explain the kinematic trends of these stellar populations (e.g. \citealt{Shenetal2010, Kunderetal2012, DiMatteoetal2015, Gomezetal2018}). On the other hand, as we will show in a subsequent paper of this series (Di Matteo et al., in prep.), the kinematic properties of the MW bulge stellar populations are well reproduced by a composite disc model. 
%The model presented here explains naturally a large fraction of the MW bulge mass, by invoking the disc populations that we already know are present in the inner Milky Way. 
These results point to the disc origin of the MW bulge and the important contribution of the chemically defined thick disc in the inner regions of the Milky Way, which of course hints at the Galaxy's formation history.
% importance of understanding the formation mechanisms of thick discs.

%---------------------------------------------------------------------------------
\begin{acknowledgements}
The authors would like to acknowledge the generous hospitality of Valerie de Lapparent and the \emph{Galaxies} team at the Institut d'Astrophysique de Paris, where this work has been partially developed. FF thanks Michael Hayden for helpful discussions on the selection function of the APOGEE survey, and Alvaro Rojas-Arriagada for useful discussions regarding the Gaia-ESO survey. PDM thanks Vanessa Hill for enriching comments on aspects of this work. This work has been supported by the ANR (Agence Nationale de la Recherche) through the MOD4Gaia project (ANR-15-CE31-0007, P.I.: P. Di Matteo). FF is supported by a postdoctoral grant from the Centre National d'Etudes Spatiales (CNES). This work was granted access to the HPC resources of CINES under the allocation A0020410154 made by GENCI. \\

Funding for the Sloan Digital Sky Survey IV has been provided by the Alfred P. Sloan Foundation, the U.S. Department of Energy Office of Science, and the Participating Institutions. SDSS-IV acknowledges
support and resources from the Center for High-Performance Computing at
the University of Utah. The SDSS web site is www.sdss.org.

SDSS-IV is managed by the Astrophysical Research Consortium for the 
Participating Institutions of the SDSS Collaboration including the 
Brazilian Participation Group, the Carnegie Institution for Science, 
Carnegie Mellon University, the Chilean Participation Group, the French Participation Group, Harvard-Smithsonian Center for Astrophysics, 
Instituto de Astrof\'isica de Canarias, The Johns Hopkins University, 
Kavli Institute for the Physics and Mathematics of the Universe (IPMU) / 
University of Tokyo, Lawrence Berkeley National Laboratory, 
Leibniz Institut f\"ur Astrophysik Potsdam (AIP),  
Max-Planck-Institut f\"ur Astronomie (MPIA Heidelberg), 
Max-Planck-Institut f\"ur Astrophysik (MPA Garching), 
Max-Planck-Institut f\"ur Extraterrestrische Physik (MPE), 
National Astronomical Observatories of China, New Mexico State University, 
New York University, University of Notre Dame, 
Observat\'ario Nacional / MCTI, The Ohio State University, 
Pennsylvania State University, Shanghai Astronomical Observatory, 
United Kingdom Participation Group,
Universidad Nacional Aut\'onoma de M\'exico, University of Arizona, 
University of Colorado Boulder, University of Oxford, University of Portsmouth, 
University of Utah, University of Virginia, University of Washington, University of Wisconsin, 
Vanderbilt University, and Yale University.
\end{acknowledgements}

%---------------------------------------------------------------------------------
\bibliographystyle{aa}
\bibliography{References}%

\begin{thebibliography}{81}
\expandafter\ifx\csname natexlab\endcsname\relax\def\natexlab#1{#1}\fi

\bibitem[{{Anders} {et~al.}(2017){Anders}, {Chiappini}, {Minchev}, {Miglio},
  {Montalb{\'a}n}, {Mosser}, {Rodrigues}, {Santiago}, {Baudin}, {Beers}, {da
  Costa}, {Garc{\'{\i}}a}, {Garc{\'{\i}}a-Hern{\'a}ndez}, {Holtzman}, {Maia},
  {Majewski}, {Mathur}, {Noels-Grotsch}, {Pan}, {Schneider}, {Schultheis},
  {Steinmetz}, {Valentini}, \& {Zamora}}]{Andersetal2017}
{Anders}, F., {Chiappini}, C., {Minchev}, I., {et~al.} 2017, \aap, 600, A70

\bibitem[{{Athanassoula}(2005)}]{Athanassoula2005}
{Athanassoula}, E. 2005, \mnras, 358, 1477

\bibitem[{{Athanassoula} {et~al.}(2016){Athanassoula}, {Rodionov}, {Peschken},
  \& {Lambert}}]{Athanassoulaetal2016}
{Athanassoula}, E., {Rodionov}, S.~A., {Peschken}, N., \& {Lambert}, J.~C.
  2016, \apj, 821, 90

\bibitem[{{Athanassoula} {et~al.}(2017){Athanassoula}, {Rodionov}, \&
  {Prantzos}}]{Athanassoulaetal2017}
{Athanassoula}, E., {Rodionov}, S.~A., \& {Prantzos}, N. 2017, \mnras, 467, L46

\bibitem[{{Bekki} \& {Tsujimoto}(2011)}]{BekkiTsujimoto2011}
{Bekki}, K. \& {Tsujimoto}, T. 2011, \mnras, 416, L60

\bibitem[{{Bensby} {et~al.}(2011){Bensby}, {Alves-Brito}, {Oey}, {Yong}, \&
  {Mel{\'e}ndez}}]{Bensbyetal2011}
{Bensby}, T., {Alves-Brito}, A., {Oey}, M.~S., {Yong}, D., \& {Mel{\'e}ndez},
  J. 2011, \apjl, 735, L46

\bibitem[{{Bensby} {et~al.}(2014){Bensby}, {Feltzing}, \&
  {Oey}}]{Bensbyetal2014}
{Bensby}, T., {Feltzing}, S., \& {Oey}, M.~S. 2014, \aap, 562, A71

\bibitem[{Binney \& Tremaine(2008)}]{BT2008}
Binney, J. \& Tremaine, S. 2008, Galactic Dynamics, 2nd edn. (Princeton
  University Press), 920

\bibitem[{{Bird} {et~al.}(2013){Bird}, {Kazantzidis}, {Weinberg}, {Guedes},
  {Callegari}, {Mayer}, \& {Madau}}]{Birdetal2013}
{Bird}, J.~C., {Kazantzidis}, S., {Weinberg}, D.~H., {et~al.} 2013, \apj, 773,
  43

\bibitem[{{Bland-Hawthorn} \& {Gerhard}(2016)}]{BlandHawthornGerhard2016}
{Bland-Hawthorn}, J. \& {Gerhard}, O. 2016, \araa, 54, 529

\bibitem[{{Bournaud} {et~al.}(2009){Bournaud}, {Elmegreen}, \&
  {Martig}}]{Bournaudetal2009}
{Bournaud}, F., {Elmegreen}, B.~G., \& {Martig}, M. 2009, \apjl, 707, L1

\bibitem[{{Bovy} {et~al.}(2012){Bovy}, {Rix}, {Liu}, {Hogg}, {Beers}, \&
  {Lee}}]{Bovyetal2012}
{Bovy}, J., {Rix}, H.-W., {Liu}, C., {et~al.} 2012, \apj, 753, 148

\bibitem[{{Bovy} {et~al.}(2016){Bovy}, {Rix}, {Schlafly}, {Nidever},
  {Holtzman}, {Shetrone}, \& {Beers}}]{Bovyetal2016}
{Bovy}, J., {Rix}, H.-W., {Schlafly}, E.~F., {et~al.} 2016, \apj, 823, 30

\bibitem[{{Brook} {et~al.}(2004){Brook}, {Kawata}, {Gibson}, \&
  {Freeman}}]{Brooketal2004}
{Brook}, C.~B., {Kawata}, D., {Gibson}, B.~K., \& {Freeman}, K.~C. 2004, \apj,
  612, 894

\bibitem[{{Calura} {et~al.}(2012){Calura}, {Gibson}, {Michel-Dansac},
  {Stinson}, {Cignoni}, {Dotter}, {Pilkington}, {House}, {Brook}, {Few},
  {Bailin}, {Couchman}, \& {Wadsley}}]{Caluraetal2012}
{Calura}, F., {Gibson}, B.~K., {Michel-Dansac}, L., {et~al.} 2012, \mnras, 427,
  1401

\bibitem[{{Combes} {et~al.}(1990){Combes}, {Debbasch}, {Friedli}, \&
  {Pfenniger}}]{Combesetal1990}
{Combes}, F., {Debbasch}, F., {Friedli}, D., \& {Pfenniger}, D. 1990, \aap,
  233, 82

\bibitem[{{Combes} \& {Sanders}(1981)}]{CombesSanders1981}
{Combes}, F. \& {Sanders}, R.~H. 1981, \aap, 96, 164

\bibitem[{{Debattista} {et~al.}(2017){Debattista}, {Ness}, {Gonzalez},
  {Freeman}, {Zoccali}, \& {Minniti}}]{Debattistaetal2017}
{Debattista}, V.~P., {Ness}, M., {Gonzalez}, O.~A., {et~al.} 2017, \mnras, 469,
  1587

\bibitem[{{Dehnen}(2000)}]{Dehnen2000}
{Dehnen}, W. 2000, \aj, 119, 800

\bibitem[{{Di Matteo}(2016)}]{DiMatteo2016}
{Di Matteo}, P. 2016, \pasa, 33, e027

\bibitem[{{Di Matteo} {et~al.}(2015){Di Matteo}, {G{\'o}mez}, {Haywood},
  {Combes}, {Lehnert}, {Ness}, {Snaith}, {Katz}, \&
  {Semelin}}]{DiMatteoetal2015}
{Di Matteo}, P., {G{\'o}mez}, A., {Haywood}, M., {et~al.} 2015, \aap, 577, A1

\bibitem[{{Di Matteo} {et~al.}(2014){Di Matteo}, {Haywood}, {G{\'o}mez}, {van
  Damme}, {Combes}, {Hall{\'e}}, {Semelin}, {Lehnert}, \&
  {Katz}}]{DiMatteoetal2014}
{Di Matteo}, P., {Haywood}, M., {G{\'o}mez}, A., {et~al.} 2014, \aap, 567, A122

\bibitem[{{Dwek} {et~al.}(1995){Dwek}, {Arendt}, {Hauser}, {Kelsall}, {Lisse},
  {Moseley}, {Silverberg}, {Sodroski}, \& {Weiland}}]{Dweketal1995}
{Dwek}, E., {Arendt}, R.~G., {Hauser}, M.~G., {et~al.} 1995, \apj, 445, 716

\bibitem[{{Eggen} {et~al.}(1962){Eggen}, {Lynden-Bell}, \&
  {Sandage}}]{Eggenetal1962}
{Eggen}, O.~J., {Lynden-Bell}, D., \& {Sandage}, A.~R. 1962, \apj, 136, 748

\bibitem[{{Fragkoudi} {et~al.}(2017{\natexlab{a}}){Fragkoudi}, {Di Matteo},
  {Haywood}, {G{\'o}mez}, {Combes}, {Katz}, \& {Semelin}}]{Fragkoudietal2017}
{Fragkoudi}, F., {Di Matteo}, P., {Haywood}, M., {et~al.} 2017{\natexlab{a}},
  \aap, 606, A47

\bibitem[{{Fragkoudi} {et~al.}(2017{\natexlab{b}}){Fragkoudi}, {Di Matteo},
  {Haywood}, {Khoperskov}, {Gomez}, {Schultheis}, {Combes}, \&
  {Semelin}}]{Fragkoudietal2017b}
{Fragkoudi}, F., {Di Matteo}, P., {Haywood}, M., {et~al.} 2017{\natexlab{b}},
  \aap, 607, L4

\bibitem[{{Freeman} {et~al.}(2013){Freeman}, {Ness}, {Wylie-de-Boer},
  {Athanassoula}, {Bland-Hawthorn}, {Asplund}, {Lewis}, {Yong}, {Lane}, {Kiss},
  \& {Ibata}}]{Freemanetal2013}
{Freeman}, K., {Ness}, M., {Wylie-de-Boer}, E., {et~al.} 2013, \mnras, 428,
  3660

\bibitem[{{Garcia Perez} {et~al.}(2017){Garcia Perez}, {Ness}, {Robin},
  {Martinez-Valpuesta}, {Sobeck}, {Zasowski}, {Majewski}, {Bovy}, {Allende
  Prieto}, {Cunha}, {Girardi}, {M{\'e}sz{\'a}ros}, {Nidever}, {Schiavon},
  {Schultheis}, {Shetrone}, \& {Smith}}]{GarciaPerrezetal2017}
{Garcia Perez}, A.~E., {Ness}, M., {Robin}, A.~C., {et~al.} 2017, ArXiv
  e-prints [\eprint[arXiv]{1712.01297}]

\bibitem[{{Gomez} {et~al.}(2018){Gomez}, {Di Matteo}, {Schultheis},
  {Fragkoudi}, {Haywood}, \& {Combes}}]{Gomezetal2018}
{Gomez}, A., {Di Matteo}, P., {Schultheis}, M., {et~al.} 2018, ArXiv e-prints
  [\eprint[arXiv]{1803.09626}]

\bibitem[{{G{\'o}mez} {et~al.}(2016){G{\'o}mez}, {Di Matteo}, {Stefanovitch},
  {Haywood}, {Combes}, {Katz}, \& {Babusiaux}}]{Gomezetal2016}
{G{\'o}mez}, A., {Di Matteo}, P., {Stefanovitch}, N., {et~al.} 2016, \aap, 589,
  A122

\bibitem[{{Gonzalez} {et~al.}(2017){Gonzalez}, {Debattista}, {Ness}, {Erwin},
  \& {Gadotti}}]{Gonzalezetal2017}
{Gonzalez}, O.~A., {Debattista}, V.~P., {Ness}, M., {Erwin}, P., \& {Gadotti},
  D.~A. 2017, \mnras, 466, L93

\bibitem[{{Gonzalez} {et~al.}(2013){Gonzalez}, {Rejkuba}, {Zoccali}, {Valent},
  {Minniti}, \& {Tobar}}]{Gonzalezetal2013}
{Gonzalez}, O.~A., {Rejkuba}, M., {Zoccali}, M., {et~al.} 2013, \aap, 552, A110

\bibitem[{{Grand} {et~al.}(2018){Grand}, {Bustamante}, {G{\'o}mez}, {Kawata},
  {Marinacci}, {Pakmor}, {Rix}, {Simpson}, {Sparre}, \&
  {Springel}}]{Grandetal2018}
{Grand}, R.~J.~J., {Bustamante}, S., {G{\'o}mez}, F.~A., {et~al.} 2018, \mnras,
  474, 3629

\bibitem[{{Grand} {et~al.}(2016){Grand}, {Springel}, {G{\'o}mez}, {Marinacci},
  {Pakmor}, {Campbell}, \& {Jenkins}}]{Grandetal2016}
{Grand}, R.~J.~J., {Springel}, V., {G{\'o}mez}, F.~A., {et~al.} 2016, \mnras,
  459, 199

\bibitem[{{Halle} {et~al.}(2015){Halle}, {Di Matteo}, {Haywood}, \&
  {Combes}}]{Halleetal2015}
{Halle}, A., {Di Matteo}, P., {Haywood}, M., \& {Combes}, F. 2015, \aap, 578,
  A58

\bibitem[{{Hayden} {et~al.}(2015){Hayden}, {Bovy}, {Holtzman}, {Nidever},
  {Bird}, {Weinberg}, {Andrews}, {Majewski}, {Allende Prieto}, {Anders},
  {Beers}, {Bizyaev}, {Chiappini}, {Cunha}, {Frinchaboy},
  {Garc{\'{\i}}a-Her{\'n}andez}, {Garc{\'{\i}}a P{\'e}rez}, {Girardi},
  {Harding}, {Hearty}, {Johnson}, {M{\'e}sz{\'a}ros}, {Minchev}, {O'Connell},
  {Pan}, {Robin}, {Schiavon}, {Schneider}, {Schultheis}, {Shetrone},
  {Skrutskie}, {Steinmetz}, {Smith}, {Wilson}, {Zamora}, \&
  {Zasowski}}]{Haydenetal2015}
{Hayden}, M.~R., {Bovy}, J., {Holtzman}, J.~A., {et~al.} 2015, \apj, 808, 132

\bibitem[{{Haywood} {et~al.}(2013){Haywood}, {Di Matteo}, {Lehnert}, {Katz}, \&
  {G{\'o}mez}}]{Haywoodetal2013}
{Haywood}, M., {Di Matteo}, P., {Lehnert}, M.~D., {Katz}, D., \& {G{\'o}mez},
  A. 2013, \aap, 560, A109

\bibitem[{{Haywood} {et~al.}(2015){Haywood}, {Di Matteo}, {Snaith}, \&
  {Lehnert}}]{Haywoodetal2015}
{Haywood}, M., {Di Matteo}, P., {Snaith}, O., \& {Lehnert}, M.~D. 2015, \aap,
  579, A5

\bibitem[{{Haywood} {et~al.}(2016){Haywood}, {Lehnert}, {Di Matteo}, {Snaith},
  {Schultheis}, {Katz}, \& {G{\'o}mez}}]{Haywoodetal2016}
{Haywood}, M., {Lehnert}, M.~D., {Di Matteo}, P., {et~al.} 2016, \aap, 589, A66

\bibitem[{{Hill} {et~al.}(2011){Hill}, {Lecureur}, {G{\'o}mez}, {Zoccali},
  {Schultheis}, {Babusiaux}, {Royer}, {Barbuy}, {Arenou}, {Minniti}, \&
  {Ortolani}}]{Hilletal2011}
{Hill}, V., {Lecureur}, A., {G{\'o}mez}, A., {et~al.} 2011, \aap, 534, A80

\bibitem[{{Johnson} {et~al.}(2011){Johnson}, {Rich}, {Fulbright}, {Valenti}, \&
  {McWilliam}}]{Johnsonetal2011}
{Johnson}, C.~I., {Rich}, R.~M., {Fulbright}, J.~P., {Valenti}, E., \&
  {McWilliam}, A. 2011, \apj, 732, 108

\bibitem[{{Kunder} {et~al.}(2012){Kunder}, {Koch}, {Rich}, {de Propris},
  {Howard}, {Stubbs}, {Johnson}, {Shen}, {Wang}, {Robin}, {Kormendy}, {Soto},
  {Frinchaboy}, {Reitzel}, {Zhao}, \& {Origlia}}]{Kunderetal2012}
{Kunder}, A., {Koch}, A., {Rich}, R.~M., {et~al.} 2012, \aj, 143, 57

\bibitem[{{Lehnert} {et~al.}(2014){Lehnert}, {Di Matteo}, {Haywood}, \&
  {Snaith}}]{Lehnertetal2014}
{Lehnert}, M.~D., {Di Matteo}, P., {Haywood}, M., \& {Snaith}, O.~N. 2014,
  \apjl, 789, L30

\bibitem[{{Lynden-Bell} \& {Kalnajs}(1972)}]{LyndenBellKalnajs1972}
{Lynden-Bell}, D. \& {Kalnajs}, A.~J. 1972, \mnras, 157, 1

\bibitem[{{Ma} {et~al.}(2017){Ma}, {Hopkins}, {Feldmann}, {Torrey},
  {Faucher-Gigu{\`e}re}, \& {Kere{\v s}}}]{Maetal2017}
{Ma}, X., {Hopkins}, P.~F., {Feldmann}, R., {et~al.} 2017, \mnras, 466, 4780

\bibitem[{{Mackereth} {et~al.}(2017){Mackereth}, {Bovy}, {Schiavon},
  {Zasowski}, {Cunha}, {Frinchaboy}, {Garc{\'{\i}}a Perez}, {Hayden},
  {Holtzman}, {Majewski}, {M{\'e}sz{\'a}ros}, {Nidever}, {Pinsonneault}, \&
  {Shetrone}}]{Mackerethetal2017}
{Mackereth}, J.~T., {Bovy}, J., {Schiavon}, R.~P., {et~al.} 2017, \mnras, 471,
  3057

\bibitem[{{Majewski} {et~al.}(2017){Majewski}, {Schiavon}, {Frinchaboy},
  {Allende Prieto}, {Barkhouser}, {Bizyaev}, {Blank}, {Brunner}, {Burton},
  {Carrera}, {Chojnowski}, {Cunha}, {Epstein}, {Fitzgerald}, {Garc{\'{\i}}a
  P{\'e}rez}, {Hearty}, {Henderson}, {Holtzman}, {Johnson}, {Lam}, {Lawler},
  {Maseman}, {M{\'e}sz{\'a}ros}, {Nelson}, {Nguyen}, {Nidever}, {Pinsonneault},
  {Shetrone}, {Smee}, {Smith}, {Stolberg}, {Skrutskie}, {Walker}, {Wilson},
  {Zasowski}, {Anders}, {Basu}, {Beland}, {Blanton}, {Bovy}, {Brownstein},
  {Carlberg}, {Chaplin}, {Chiappini}, {Eisenstein}, {Elsworth}, {Feuillet},
  {Fleming}, {Galbraith-Frew}, {Garc{\'{\i}}a}, {Garc{\'{\i}}a-Hern{\'a}ndez},
  {Gillespie}, {Girardi}, {Gunn}, {Hasselquist}, {Hayden}, {Hekker}, {Ivans},
  {Kinemuchi}, {Klaene}, {Mahadevan}, {Mathur}, {Mosser}, {Muna}, {Munn},
  {Nichol}, {O'Connell}, {Parejko}, {Robin}, {Rocha-Pinto}, {Schultheis},
  {Serenelli}, {Shane}, {Silva Aguirre}, {Sobeck}, {Thompson}, {Troup},
  {Weinberg}, \& {Zamora}}]{Majewskietal2017}
{Majewski}, S.~R., {Schiavon}, R.~P., {Frinchaboy}, P.~M., {et~al.} 2017, \aj,
  154, 94

\bibitem[{{Martig} {et~al.}(2014){Martig}, {Minchev}, \&
  {Flynn}}]{Martigetal2014}
{Martig}, M., {Minchev}, I., \& {Flynn}, C. 2014, \mnras, 442, 2474

\bibitem[{{Martinez-Valpuesta} \&
  {Gerhard}(2013)}]{MartinezValpuestaGerhard2013}
{Martinez-Valpuesta}, I. \& {Gerhard}, O. 2013, \apjl, 766, L3

\bibitem[{{Martinez-Valpuesta} {et~al.}(2006){Martinez-Valpuesta}, {Shlosman},
  \& {Heller}}]{MartinezValpuestaetal2006}
{Martinez-Valpuesta}, I., {Shlosman}, I., \& {Heller}, C. 2006, \apj, 637, 214

\bibitem[{{McWilliam} \& {Rich}(1994)}]{McWilliamRich1994}
{McWilliam}, A. \& {Rich}, R.~M. 1994, \apjs, 91, 749

\bibitem[{{McWilliam} \& {Zoccali}(2010)}]{McWilliamandZoccali2010}
{McWilliam}, A. \& {Zoccali}, M. 2010, \apj, 724, 1491

\bibitem[{{Minchev} {et~al.}(2015){Minchev}, {Martig}, {Streich},
  {Scannapieco}, {de Jong}, \& {Steinmetz}}]{Minchevetal2015}
{Minchev}, I., {Martig}, M., {Streich}, D., {et~al.} 2015, \apjl, 804, L9

\bibitem[{{Minniti} {et~al.}(1995){Minniti}, {Olszewski}, {Liebert}, {White},
  {Hill}, \& {Irwin}}]{Minnitietal1995}
{Minniti}, D., {Olszewski}, E.~W., {Liebert}, J., {et~al.} 1995, \mnras, 277,
  1293

\bibitem[{{Nandakumar} {et~al.}(2017){Nandakumar}, {Schultheis}, {Hayden},
  {Rojas-Arriagada}, {Kordopatis}, \& {Haywood}}]{Nandakumaretal2017}
{Nandakumar}, G., {Schultheis}, M., {Hayden}, M., {et~al.} 2017, \aap, 606, A97

\bibitem[{{Nataf} {et~al.}(2010){Nataf}, {Udalski}, {Gould}, {Fouqu{\'e}}, \&
  {Stanek}}]{Natafetal2010}
{Nataf}, D.~M., {Udalski}, A., {Gould}, A., {Fouqu{\'e}}, P., \& {Stanek},
  K.~Z. 2010, \apjl, 721, L28

\bibitem[{{Ness} \& {Freeman}(2016)}]{NessFreeman2016}
{Ness}, M. \& {Freeman}, K. 2016, \pasa, 33, e022

\bibitem[{{Ness} {et~al.}(2013){Ness}, {Freeman}, {Athanassoula},
  {Wylie-de-Boer}, {Bland-Hawthorn}, {Asplund}, {Lewis}, {Yong}, {Lane}, \&
  {Kiss}}]{Nessetal2013a}
{Ness}, M., {Freeman}, K., {Athanassoula}, E., {et~al.} 2013, \mnras, 430, 836

\bibitem[{{Ness} \& {Lang}(2016)}]{NessLang2016}
{Ness}, M. \& {Lang}, D. 2016, \aj, 152, 14

\bibitem[{{Ness} {et~al.}(2012)}]{Nessetal2012}
{Ness}, M. {et~al.} 2012, \apj, 756, 22

\bibitem[{{Obreja} {et~al.}(2013){Obreja}, {Dom{\'{\i}}nguez-Tenreiro},
  {Brook}, {Mart{\'{\i}}nez-Serrano}, {Dom{\'e}nech-Moral}, {Serna},
  {Moll{\'a}}, \& {Stinson}}]{Obrejaetal2013}
{Obreja}, A., {Dom{\'{\i}}nguez-Tenreiro}, R., {Brook}, C., {et~al.} 2013,
  \apj, 763, 26

\bibitem[{{P{\'e}rez-Villegas} {et~al.}(2017){P{\'e}rez-Villegas}, {Portail},
  {Wegg}, \& {Gerhard}}]{PerezVillegasetal2017}
{P{\'e}rez-Villegas}, A., {Portail}, M., {Wegg}, C., \& {Gerhard}, O. 2017,
  \apjl, 840, L2

\bibitem[{{Portail} {et~al.}(2017){Portail}, {Wegg}, {Gerhard}, \&
  {Ness}}]{Portailetal2017}
{Portail}, M., {Wegg}, C., {Gerhard}, O., \& {Ness}, M. 2017, \mnras, 470, 1233

\bibitem[{{Quillen} {et~al.}(2014){Quillen}, {Minchev}, {Sharma}, {Qin}, \& {Di
  Matteo}}]{Quillenetal2014}
{Quillen}, A.~C., {Minchev}, I., {Sharma}, S., {Qin}, Y.-J., \& {Di Matteo}, P.
  2014, \mnras, 437, 1284

\bibitem[{{Raha} {et~al.}(1991){Raha}, {Sellwood}, {James}, \&
  {Kahn}}]{Rahaetal1991}
{Raha}, N., {Sellwood}, J.~A., {James}, R.~A., \& {Kahn}, F.~D. 1991, \nat,
  352, 411

\bibitem[{{Rich} {et~al.}(2007){Rich}, {Origlia}, \& {Valenti}}]{Richetal2007}
{Rich}, R.~M., {Origlia}, L., \& {Valenti}, E. 2007, \apjl, 665, L119

\bibitem[{{Rodionov} {et~al.}(2009){Rodionov}, {Athanassoula}, \&
  {Sotnikova}}]{Rodionovetal2009}
{Rodionov}, S.~A., {Athanassoula}, E., \& {Sotnikova}, N.~Y. 2009, \mnras, 392,
  904

\bibitem[{{Rojas-Arriagada} {et~al.}(2014){Rojas-Arriagada}, {Recio-Blanco},
  {Hill}, {de Laverny}, {Schultheis}, {Babusiaux}, {Zoccali}, {Minniti},
  {Gonzalez}, {Feltzing}, {Gilmore}, {Randich}, {Vallenari},
  {et~al.}}]{RojasArriagadaetal2014}
{Rojas-Arriagada}, A., {Recio-Blanco}, A., {Hill}, V., {et~al.} 2014, \aap,
  569, A103

\bibitem[{{Schultheis} {et~al.}(2017){Schultheis}, {Rojas-Arriagada},
  {Garc{\'{\i}}a P{\'e}rez}, {J{\"o}nsson}, {Hayden}, {Nandakumar}, {Cunha},
  {Allende Prieto}, {Holtzman}, {Beers}, {Bizyaev}, {Brinkmann}, {Carrera},
  {Cohen}, {Geisler}, {Hearty}, {Fernandez-Tricado}, {Maraston}, {Minnitti},
  {Nitschelm}, {Roman-Lopes}, {Schneider}, {Tang}, {Villanova}, {Zasowski}, \&
  {Majewski}}]{Schultheisetal2017}
{Schultheis}, M., {Rojas-Arriagada}, A., {Garc{\'{\i}}a P{\'e}rez}, A.~E.,
  {et~al.} 2017, \aap, 600, A14

\bibitem[{{SDSS Collaboration} {et~al.}(2016){SDSS Collaboration}, {Albareti},
  {Allende Prieto}, {Almeida}, {Anders}, {Anderson}, {Andrews},
  {Aragon-Salamanca}, {Argudo-Fernandez}, {Armengaud}, \& et~al.}]{SDSSDR13}
{SDSS Collaboration}, {Albareti}, F.~D., {Allende Prieto}, C., {et~al.} 2016,
  ArXiv e-prints [\eprint[arXiv]{1608.02013}]

\bibitem[{{Shen} {et~al.}(2010){Shen}, {Rich}, {Kormendy}, {Howard}, {De
  Propris}, \& {Kunder}}]{Shenetal2010}
{Shen}, J., {Rich}, R.~M., {Kormendy}, J., {et~al.} 2010, \apjl, 720, L72

\bibitem[{{Snaith} {et~al.}(2015){Snaith}, {Haywood}, {Di Matteo}, {Lehnert},
  {Combes}, {Katz}, \& {G{\'o}mez}}]{Snaithetal2015}
{Snaith}, O., {Haywood}, M., {Di Matteo}, P., {et~al.} 2015, \aap, 578, A87

\bibitem[{{Snaith} {et~al.}(2014){Snaith}, {Haywood}, {Di Matteo}, {Lehnert},
  {Combes}, {Katz}, \& {G{\'o}mez}}]{Snaithetal2014}
{Snaith}, O.~N., {Haywood}, M., {Di Matteo}, P., {et~al.} 2014, \apjl, 781, L31

\bibitem[{{Tissera} {et~al.}(2018){Tissera}, {Machado}, {Carollo}, {Minniti},
  {Beers}, {Zoccali}, \& {Meza}}]{Tisseraetal2018}
{Tissera}, P.~B., {Machado}, R.~E.~G., {Carollo}, D., {et~al.} 2018, \mnras,
  473, 1656

\bibitem[{{Wang} {et~al.}(2016){Wang}, {Shi}, {Pan}, {Chen}, {Zhao}, \&
  {Wicker}}]{Wangetal2016}
{Wang}, J., {Shi}, J., {Pan}, K., {et~al.} 2016, \mnras, 460, 3179

\bibitem[{{Wegg} \& {Gerhard}(2013)}]{WeggGerhard2013}
{Wegg}, C. \& {Gerhard}, O. 2013, \mnras, 435, 1874

\bibitem[{{Wisnioski} {et~al.}(2015)}]{Wisnioskietal2015}
{Wisnioski}, E. {et~al.} 2015, \apj, 799, 209

\bibitem[{{Wuyts} {et~al.}(2016)}]{Wuytsetal2016}
{Wuyts}, E. {et~al.} 2016, \apj, 827, 74

\bibitem[{{Zoccali} {et~al.}(2014){Zoccali}, {Gonzalez}, {Vasquez}, {Hill},
  {Rejkuba}, {Valenti}, {Renzini}, {Rojas-Arriagada}, {Martinez-Valpuesta},
  {Babusiaux}, {Brown}, {Minniti}, \& {McWilliam}}]{Zoccalietal2014}
{Zoccali}, M., {Gonzalez}, O.~A., {Vasquez}, S., {et~al.} 2014, \aap, 562, A66

\bibitem[{{Zoccali} {et~al.}(2008){Zoccali}, {Hill}, {Lecureur}, {Barbuy},
  {Renzini}, {Minniti}, {G{\'o}mez}, \& {Ortolani}}]{Zoccalietal2008}
{Zoccali}, M., {Hill}, V., {Lecureur}, A., {et~al.} 2008, \aap, 486, 177

\bibitem[{{Zoccali} {et~al.}(2017){Zoccali}, {Vasquez}, {Gonzalez}, {Valenti},
  {Rojas-Arriagada}, {Minniti}, {Rejkuba}, {Minniti}, {McWilliam}, {Babusiaux},
  {Hill}, \& {Renzini}}]{Zoccalietal2017}
{Zoccali}, M., {Vasquez}, S., {Gonzalez}, O.~A., {et~al.} 2017, \aap, 599, A12

\end{thebibliography}
%\end{thebibliography}

\end{document}